\documentclass[pra,twocolumn,showkeys,showpacs,10pt,superscriptaddress,aps]{revtex4-2}
\usepackage{amsmath,amssymb,amsthm}
\usepackage[english]{babel}
\addto\captionsenglish{}
\usepackage{dcolumn}
\usepackage{mathtools}
\usepackage{hyperref}
\usepackage[svgnames]{xcolor}  
\hypersetup{
    colorlinks,
    linkcolor={red!50!black},
    citecolor={blue!50!black},
    urlcolor={blue!80!black}
}
\usepackage{physics}
\usepackage{dsfont}
\usepackage{slashed}
\usepackage{mathrsfs}

\usepackage{float}

\usepackage[T1]{fontenc}
\usepackage{bm}

\usepackage{textcomp}
\usepackage{graphicx} 
\usepackage[utf8]{inputenc}
\usepackage[normal, FIGBOTCAP]{subfigure}
\usepackage{tikz}
\usepackage{mathtools}

\usepackage{pgf}
\usepackage{verbatim} 
\usepackage{import}
\usepackage[separate-uncertainty = true,multi-part-units=brackets]{siunitx}

\usepackage{units}
\definecolor{tud2a}{RGB}{0 156 218}
\definecolor{tud2b}{RGB}{0 131 204}
\definecolor{tud2d}{RGB}{0 78 115}
\definecolor{tud9b}{RGB}{230 0 26}
\definecolor{tud9c}{RGB}{185 15 34}
\definecolor{tud9d}{RGB}{156 28 38}
\definecolor{tud11b}{RGB}{114 16 133}
\definecolor{greena}{HTML}{58C25D}
\definecolor{greenb}{HTML}{36A63B}
\definecolor{greenc}{HTML}{337B37}
\definecolor{greend}{HTML}{195C29}
\definecolor{verylightgray}{rgb}{0.87, 0.87, 0.87}

\usepackage{latexsym}
\usepackage{wasysym}  
\newcommand{\floor}[1]{\left\lfloor #1 \right\rfloor}
\newcommand{\ceil}[1]{\left\lceil #1 \right\rceil}
\newcommand{\rfr}{\vec{\mathfrak{r}}(t)}
\newcommand{\pfr}{\vec{\mathfrak{p}}}
\newcommand{\ropv}{\op{\vec{r}}}
\newcommand{\rv}{\vec{r}}
\newcommand{\kv}{\vec{k}}

\newcommand{\kl}{k_L}
\newcommand{\klvec}{\vec{k}_L}
\newcommand{\wrec}{\omega_r}
\newcommand{\wrecd}{\omega_{2r}}
\newcommand{\mass}{M}

\newcommand{\omph}{\Omega}
\newcommand{\omphr}{\omph_{r}}

\newcommand{\ompheffdelta}{\Omega_{\dk}}
\newcommand{\tomph}{\tilde{\Omega}}
\newcommand{\tOmega}{\tilde{\Omega}}
\newcommand{\bomph}{\mathcal{I}}

\newcommand{\xr}{x_R}
\newcommand{\wa}{\omega_0}
\newcommand{\wl}{\omega_L}

\newcommand{\dk}{\kappa}

\newcommand{\bk}{\mathbf{k}}
\newcommand{\bq}{\mathbf{q}}

\newcommand{\df}{\mbox{$\delta\omega$}}
\newcommand{\dr}{\mbox{$\delta$}}
\newcommand{\tth}{t}
\newcommand{\gf}{\mathcal{G}}
\newcommand{\mfn}{N}
\newcommand{\nn}{\mathfrak{n}}
\newcommand{\gnn}{\mathfrak{n}}
\newcommand{\din}{n}

\newcommand{\spi}{\tau_{\pi}}

\newcommand{\tspi}{\tilde{\tau}_{\pi}}

\newcommand{\xavec}{\vec{r}_0}
\newcommand{\kavec}{\vec{k}_0}
\newcommand{\ravec}{\vec{r}_0}

\renewcommand{\op}[1]{\hat{#1}}
\renewcommand{\vec}[1]{\bm{#1}}

\newcommand{\nss}{\negthinspace}
\newcommand{\nms}{\negmedspace}

\newcommand{\of}[1]{\!\left(\! #1 \!\right)}

\newcommand{\oppv}{\op{\vec{p}}}
\newcommand{\popv}{\oppv}
\newcommand{\opx}{\op{x}}
\newcommand{\opp}{\op{p}}

\newcommand{\fig}{Fig.\,}
\newcommand{\infig}{in \fig}

\newcommand{\scross}{$\boldsymbol{\times}$}
\newcommand{\striangdown}{$\boldsymbol{\triangledown}$}

\newcommand{\striangleft}{$\boldsymbol{\triangleleft}$}
\newcommand{\striangright}{$\boldsymbol{\triangleright}$}
\newcommand{\ssquare}{$\boldsymbol{\square}$}
\newcommand{\scircle}{\Large{$\boldsymbol{\circ}$}}

\newcommand{\Tint}{\Delta \tau}
\newcommand{\hdkh}{\mathcal{H}}

\newcommand{\crosss}{\boldsymbol{\times}}

\newcommand{\vsp}{\vspace{-3mm}}

\newcommand{\av}[1]{\langle #1 \rangle}

\begin{document}

 \title{Aberrations in (3+1)D Bragg diffraction using pulsed Gaussian laser beams}

\newcommand{\affT}{Technical University of Darmstadt, Institute of Applied Physics, Germany}
\newcommand{\affZ}{ZARM, Universität Bremen, Germany}
\author{A. Neumann}
\email{Antje.Neumann@tu-darmstadt.de}
\affiliation{\affT}

\author{M. Gebbe}
\affiliation{\affZ}

\author{R. Walser}%
\affiliation{\affT}

\date{\today}

\begin{abstract}
We analyze the transfer function of a three-dimensional atomic Bragg beamsplitter formed by two counterpropagating pulsed Gaussian laser beams.  Even for ultracold atomic ensembles, 
the transfer efficiency depends significantly on the residual velocity of the particles as well as on losses into higher diffraction orders.  
Additional aberrations are caused by the spatial intensity variation and wavefront curvature of the Gaussian beam envelope, studied with (3+1)D numerical simulations. 
The temporal pulse shape also affects the transfer efficiency significantly. Thus, we  consider the practically important rectangular-, \mbox{Gaussian-,} Blackman- and hyperbolic secant pulses. For the latter, we can describe the  time-dependent response analytically with the Demkov-Kunike method. 
The experimentally observed stretching of the $\pi$-pulse time is explained from a renormalization of the simple Pendell\"osung frequency. 
Finally, we compare the analytical predictions for the velocity-dependent transfer function with effective (1+1)D numerical simulations for pulsed Gaussian beams, as well as experimental data and find very good agreement, considering a mixture of Bose-Einstein condensate and thermal cloud.
\end{abstract}

\keywords{ Bragg diffraction, atomic beamsplitter, atom optics, atom interferometer, Bose-Einstein condensation}

\maketitle
\section{Introduction}\label{sec:intro}\vsp
Atoms represent the ultimate ``abrasion free'' quantum sensors for electro-magnetic fields and gravitational forces. By a feat of nature, they occur with bosonic or fermionic attributes, but are produced otherwise identically without “manufacturing tolerance”.
A beamsplitter based on Bragg diffraction \cite{Kazantsev1990,Berman1997, Cronin2009, Tino2014} prepares superpositions of matter wavepackets by transferring photon momentum from a laser to an atomic 
wave. Controlling the diffracted populations, one can realize a beamsplitter and a mirror. These devices are the central component of a matter-wave interferometer
\cite{Kazantsev1990,Berman1997, Cronin2009, Tino2014, Abend2019}.
 Due to the well-defined 
 properties of the atomic test masses and their precise control by laser light, matter-wave interferometry 
can be used for high-precision measurements of rotation and acceleration. 
Applications range from tests of fundamental physics, like the equivalence principle 
 \cite{Fray2004, Schlippert2014, Zhou2015, Altschul2015, Bonnin2015, Barrett2016, Williams2016, Asenbaum2020} or quantum electrodynamics \cite{Arvanitaki2008, Bouchendira2011, Parker2018}, to inertial sensing \cite{Peters2001, McGuirk2002, Dimopoulos2007, Aguilera2014,  Dutta2016}.
Like all imaging systems, atom optics suffer from imperfections and an accurate characterization is 
required in order to rectify them. 
 This is relevant for high-precision experiments, for instance gravimetry \cite{Peters2001, Schkolnik2015, Xu2019} and extended 
  free-fall experiments in large fountains, micro-gravity and space 
 \cite{Geiger2011, Muntinga2013, Kovachy2015b, Schuldt2015, Becker2018, Elliott2018, Tino2019}. Such challenging experiments require realistic modeling and aberration studies, ideally hinting towards rectification.
  
For ultra-sensitive atom interferometry a large and precise momentum transfer is essential \cite{Chiow2011, McDonald2013,Plotkin-Swing2018b, Gebbe2019}. 
Bragg scattering of atoms from a moving standing light wave 
\cite{Martin1988, Giltner1995a, Oberthaler1996, Kunze1996}, potentially in a retroreflective geometry
\cite{Giese2013b,Hartmann2020a}, provides an efficient transfer of photon momentum without changing the atomic internal state.
In contrast, Raman scattering \cite{Kasevich1991, Berman1997} couples different atomic internal states, enabling velocity filtering \cite{Kasevich1991a,Neumann2020}.
While Raman pulses have lower demands on the atomic momentum distribution \cite{Hartmann2020a, Hartmann2020b}, Bragg pulses can be used for higher-order diffraction, also in combination with Bloch oscillations 
\cite{Dahan1996, Wilkinson1996, Peik1997, Muller2009, Clade2009, McDonald2013, Parker2018, Gebbe2019, McAlpine2020}.

The quasi-Bragg regime of atomic diffraction
with smooth temporal pulse shapes is optimal
\cite{Muller2008, Muller2009, Chiow2011, Kovachy2012, McDonald2013, Kovachy2015c,Ahlers2016b, Plotkin-Swing2018b, Parker2018, Gebbe2019}. 
It provides a high diffraction efficiency with moderate velocity selectivity for relevant pulse duration.   However, losses into higher diffraction orders and the velocity dispersion must be considered because atomic clouds do have a finite momentum width. 
 
The limit of the deep-Bragg regime
with long interaction times and 
  shallow optical potentials gives a perfect on-resonance diffraction efficiency but remains very narrow in momentum \cite{Berman1997}. However, it is suitable to generate velocity filters  
  \cite{Kovachy2015b, McDonald2013}.
 In the opposite Raman-Nath limit 
short laser pulses provide a vanishing velocity dispersion but  the diffraction efficiency is very low \cite{Berman1997}. 
 Despite their restrictions, both limits are popular as simple analytical solutions can be given for rectangular pulse shapes and plane-wave laser beams.
  
 For smooth temporal envelopes there exist models based on adiabatic elimination of the off-resonant coupled diffraction orders, solving the effective two-level dynamics \cite{Muller2008} and considering the velocity dispersion \cite{Szigeti2012, Giese2013b}. 
The Bloch-band picture is suitable in the quasi-Bragg regime for 
sufficiently slow (adiabatic) pulses \cite{Gochnauer2019}.
 An analytic theory for smooth pulses based on the adiabatic theorem for single quasi-Bragg pulses is given in \cite{Siemss2020}. Here, Doppler shifts are considered in terms of perturbation theory to take finite atomic momentum widths into account.

Besides temporal envelopes, spatial envelopes also affect the beamsplitter efficiency \cite{Muller2008b, Schkolnik2015, Kovachy2015b, Parker2018}, especially for large momentum transfer interferometers. In particular, spatial variations due to three-dimensional Gauss-Laguerre beams lead to aberrations.\\

\begin{figure*}[t]
	\includegraphics[width=2\columnwidth]{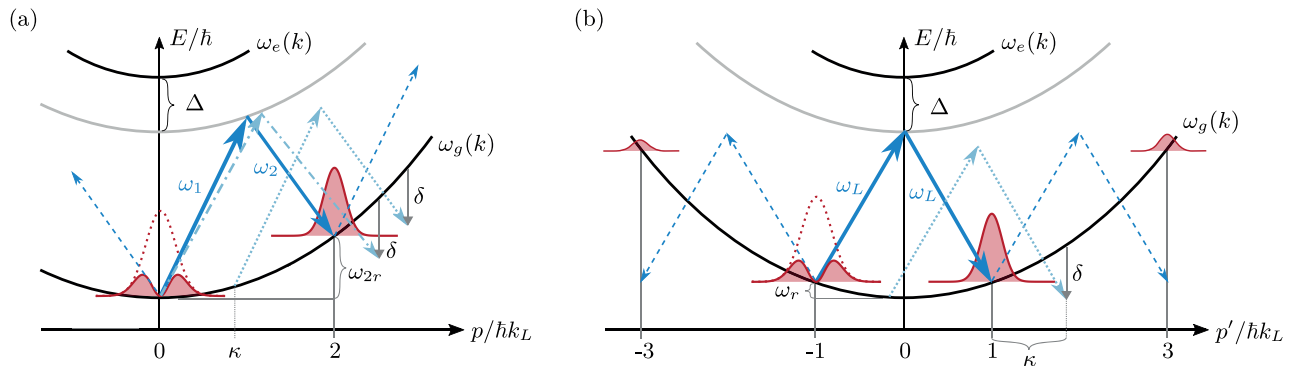}
	\vspace{-2mm}
	\caption{Bragg diffraction:
		energy diagram versus atomic wavenumber  
		$k=p/\hbar$ in units of $\kl$ \eqref{eq:kl} in the lab frame $S$ (a) 
		and an inertial frame $S'$ (b) moving with velocity $v_g$ \eqref{vgroup}. 
		Ground- and excited state eigen-frequencies of a free particle are
		  $\omega_g(k)$, $\omega_e(k)$, the two-photon and one-photon recoil frequencies $\wrecd$  and $\wrec$, respectively.
		  In frame $S$, we show that a deliberate detuning $\df$ \eqref{eq:dw} of the laser frequencies $\omega_1,\omega_2$ leads to the same fr6equency gap $\dr$ \eqref{eq:drdf} (dashed-dotted arrows), as caused by a finite initial momentum $p_i=\dk\hbar \kl$ \eqref{eq:drdk} (dotted arrows). 
	    In frame $S'$, the counterpropagating lasers have equal frequencies $\omega_{1,2}'=\omega_L$ \eqref{eq:kl} and link $p_i'=-\hbar\kl$ with $p_f'=\hbar\kl$. 
	    The velocity selectivity of Bragg scattering leads to an incomplete transfer in the momentum ensembles (red, shadowed). 
	    Odd momenta $\pm 3\kl, \pm 5\kl,\ldots$ are populated  by higher order diffraction. 
		\label{fig:braggdiffskizze}}
\end{figure*}
In this article, we will revisit atomic beamsplitters in a moving frame in 
Sec.\,\ref{sec:bs}.
We compare two common methods to solve the Schr\"odinger equation with plane-wave laser beams in Sec.~\ref{sec:pw}. 
This is the Bloch-wave ansatz and an ad-hoc ansatz, which leads to a more convenient extended zone scheme. 
In Sec.\,\ref{sec:charac}, we study aberrations
due to velocity selectivity, higher diffraction orders, spatial variations of the beam intensities, wavefront curvatures and the influence of four non-adiabatic temporal 
pulse envelopes in terms of the complex transfer function and the fidelity. 
We introduce an explicitly solvable Demkov-Kunike type model, which applies to hyperbolic sech pulses. 
With the full (3+1)D simulations the effects of spatially Gauss-Laguerre laser beams are studied.
 Finally, we gauge simulations and explicit models to experimental data in Sec.\,\ref{sec:exp}.

\section{Matter-wave Bragg beamsplitter}\label{sec:bs}\vsp
\subsection{Conservation laws}\vsp
The basic mechanism of an atomic beamsplitter is the stimulated 
absorption and emission of two photons from bichromatic, counterpropagating laser
 beams \cite{Bernhardt1981,Kazantsev1990}.
 This process is depicted \infig~\ref{fig:braggdiffskizze}a and satisfies energy and momentum conservation
 \begin{equation}\label{eq:energyconserv}
 \frac{p_i^2}{2M}+\hbar \omega_1=
\frac{p_f^2}{2M}+\hbar \omega_2, \qquad
  p_i+\hbar k_1=p_f-\hbar k_2.
 \end{equation}
  Here, $p_{i,f}$ are the initial and final momenta 
  of the particle with mass $M$, $\pm \hbar k_{1,2}$ are photon momenta and $\omega_{1,2}$ are the laser frequencies. 
We choose to work with positive wavenumbers $k_{1,2} >0$ and
emphasize the propagation directions with explicit signs, but retain the directionality of $p_{i,f}$.  Frequency and wavenumber are coupled by the vacuum dispersion relation $\omega=c k$, with the speed of light $c$. 
One chooses counterpropagating beams
to maximize the momentum transfer 
$p_f-p_i=2 \hbar \kl$, introducing the average wavenumber and frequency 
\begin{equation}\label{eq:kl}
\kl\equiv \frac{k_1+k_2}{2},\qquad
\omega_L\equiv c \kl.
\end{equation}

Wave mechanics considers superpositions of momentum states $\ket{g,p_i}$ and $\ket{g,p_f}$ in the internal atomic ground state $g$. 
For atoms initially at rest $p_i=0$, energy and momentum conservation \eqref{eq:energyconserv} requires laser frequencies
 \begin{equation}\label{eq:resfrequ}
\omega_1=\omega_2+\wrecd\approx \omega_2+\frac{\hbar (2k_2)^2}{2M}.
 \end{equation}
Due to the two-photon recoil, we need to introduce 
 \begin{equation}\label{eq:recoilfrequencies}
	\wrecd\equiv\frac{\hbar (2\kl)^2}{2M}=4 \wrec,
\end{equation} as the two-photon frequency $\wrecd$
in terms of the single photon frequency $\wrec$.
The approximation \eqref{eq:resfrequ} holds for non-relativistic energies, just as the kinetic energy in \eqref{eq:energyconserv}.
\subsection{Off-resonant response}\label{offresponse}\vsp
Releasing ultracold atomic ensembles from traps provides localized wavepackets with a finite momentum dispersion. Therefore, one needs to study the response of the Bragg beamsplitter with finite initial- and final momenta $\bar{p}_{i} = \dk\hbar\kl$,  $\bar{p}_{f}= (2+\dk)\hbar\kl$, introducing a dimensionless momentum
$\dk$. This opens a frequency gap 
\begin{equation}
	\label{eq:drdk}
	\dr \equiv 
	\frac{\bar{p}_{f}^2}{2M\hbar}
	+\omega_2
	-  \frac{\bar{p}_{i}^2}{2M\hbar}
	-\omega_1
	=
	\wrecd\dk,
\end{equation} 
shown \infig\ref{fig:braggdiffskizze} (a).

Alternatively, one can also probe the momentum response by a detuning of the laser frequencies $\tilde{\omega}_{1,2}$ from the resonant values $\omega_{1,2}$ in \eqref{eq:resfrequ}. Conveniently, this detuning is measured by
\begin{equation}\label{eq:dw}
	\df \equiv\omega_1 -\omega_2 + \tilde{\omega}_2-\tilde{\omega}_1.
\end{equation}
Dash-dotted arrows mark the deviant frequencies in Fig.~\ref{fig:braggdiffskizze} (a).
For a particle, which is initially at rest \mbox{$\tilde{p}_i=0$} and
acquires a momentum
$\tilde{p}_f=\hbar(\tilde{k}_1+\tilde{k}_2)$ 
after the momentum transfer,
one obtains a frequency gap 
\begin{equation}\label{eq:drdf}
\dr = \frac{\tilde{p}_f^2}{2M\hbar}
+\tilde{\omega}_{2}
-\tilde{\omega}_{1}\approx \df.
\end{equation} 
The approximation holds for $|\tilde{\omega}_{1,2}-\omega_{1,2}|\ll\omega_L$, which is satisfied very well in the present context.
Comparing Eqs.~\eqref{eq:drdk} and \eqref{eq:drdf}, 
one finds a linear relation  
\begin{equation}\label{eq:dwdk}
\df=\wrecd \dk,
\end{equation}
between laser-frequency mismatch $\df$ and the dimensionless initial particle momentum $\dk$. 
Therefore, both realizations are suitable to probe the momentum response of Bragg diffraction and their results are related by Eq.~\eqref{eq:dwdk}. 

Experimentally, it is advantageous to modify the laser-frequencies (cf. Sec.~\ref{sec:exp})
and to prepare atomic wavepackets initially at rest in the lab-frame $S$.
Theoretically, it is beneficial to emphasize  the symmetries of the system. Therefore, we  will
adopt a moving inertial frame $S'$, wherein the Doppler-shifted laser-frequencies coincide and
the momentum coupled states  
$p_i^\prime=-\hbar \kl$, $p_f^\prime=+\hbar \kl$ are distributed symmetrically (cf. Sec.~\ref{sec:bichromfields}, App.~\ref{app:galileo}). This is depicted \infig\ref{fig:braggdiffskizze}b.

\subsection{Counterpropagating, bichromatic fields}\label{sec:bichromfields}\vsp
The superposition of two counterpropagating laser beams 
$\vec{E}=\vec{E}_{1} + \vec{E}_{2}$, is defined by the constituent fields $ \vec{E}_i=\Re[ \vec{E}_i^{(+)}]$ with the positive frequency components 
\begin{equation}
\vec{E}_i^{(+)} (t,\rv)
=\boldsymbol{\epsilon}_i 
e^{-i \phi_i(t,x)}
\mathcal{E}_i(t,\rv).
\end{equation}
Here, $\boldsymbol{\epsilon}_i$ denote the polarization vectors,  
$\mathcal{E}_i(t,\rv)$ the slowly varying complex Gaussian envelopes and  
$\phi_1(t,x)= \omega_1 t-k_1 x$,
$\phi_2(t,x)=\omega_2 t+k_2 x$ 
are the rapidly oscillating carrier phases
for fields propagating along the x-direction \cite{Sulzbach2019}
(cf. App.\,\ref{app:galileo} and \ref{app:gaussianbeams}).
From the superposition of two scalar counterpropagating 
bichromatic fields
\begin{equation}
\mathcal{E}= e^{-i\phi_1(t,x)} \mathcal{E}_1 +
e^{-i\phi_2(t,x)} \mathcal{E}_2 ,
\end{equation} 
one obtains a steady motion of the intensity pattern 
\begin{equation}
|\mathcal{E}|^2=|\mathcal{E}_1|^2+|\mathcal{E}_2|^2+ 
2 \Re\left[\mathcal{E}_2^\ast \mathcal{E}_1^{\phantom{*}}  e^{i(k_1+k_2)(x-v_{g} t)}\right],
\end{equation} 
where nodes move with the group velocity 
\begin{equation}\label{vgroup}
v_{g}=
\frac{\omega_1-\omega_2}{\omega_1+\omega_2} c,\qquad
|v_{g}|= \frac{\wrecd}{2\omega_L}c\ll c. 
\end{equation}
If the lab frame $S$ has the coordinates $x$,  then the moving interference pattern defines
another inertial frame $S'$, where the grating is at rest
and the coordinates
\begin{equation}\label{comov}
x'=x-v_g t, 
\end{equation}
are related to the lab frame coordinates $x$ by a passive Galilean transformation. 
\subsection{Interaction energy}\vsp
The atom is represented by a ground $\ket{g}$ and an 
excited state $\ket{e}$. These levels are separated by the transition frequency 
$\wa=\omega_e-\omega_g$ and coupled by the electric dipole matrix element
$\vec{d}_{eg}=\bra{e}\hat{\vec{d}}\ket{g}$.
To neglect spontaneous emissions,
the lasers are far-detuned from the atomic resonance 
frequencies $| \wa- \omega_i|\gg\Gamma$, where $\Gamma$ is the natural linewidth of the transition.
In the lab frame $S$
the Hamilton operator of an atom with mass $\mass$ reads
\begin{align}\label{eq:ham0}
\op{H}(t) =&
\frac{\op{\vec{p}}^2}{2\mass}
+\hbar \omega_g\op{\sigma}_g +\hbar \omega_e\op{\sigma}_e+V(t,\op{\rv}),\\
V(t,\rv)=& \frac{\hbar}{2}
\op{\sigma}^\dag
\sum_{i=1}^2\Omega_i(t,\rv)e^{-i\phi_i(t,x)}+\text{h.c.},\notag
\end{align} 
using the spin operators 
$\op{\sigma}_{i=e,g}=\ketbra{i}{i}$ and $\op{\sigma} =\ketbra{g}{e}$.
Here, we  evaluate  the electric dipole interaction energy in the rotating-wave approximation and denote the Rabi frequencies as 
$\Omega_i(t,\rv)= 
-\boldsymbol{\varepsilon}_i \vdot \vec{d}_{ge}\,\mathcal{E}_i(t,\rv)/\hbar$ .

If we transform this Hamilton operator to the frame $S'$, comoving with the nodes of the interference pattern \eqref{comov}, and use a corotating internal frame \eqref{eq:sgtrafoII}, it reads
\begin{equation}
	\label{eq:Gham}
\begin{split}
\op{H}''(t) =&\frac{\op{\vec{p}}^2}{2\mass}
-\hbar\Delta\op{\sigma}_{e}+
\frac{\hbar}{2} \op{\sigma} ^\dagger
\left[\tilde{\Omega}_1(t,\op{\rv}) e^{i\kl \opx}\right.\\
&+\left. \tilde{\Omega}_2(t,\op{\rv}) e^{-i\kl \opx}\right]
+\text{h.c.}
\end{split}
\end{equation}
In this specific frame  the atom responds only to a carrier wavenumber $\kl$. 
We measure the laser detuning $\Delta \equiv \wl-\wa$ with respect to the common Doppler-shifted frequency $\omega_L$. The Rabi frequencies $\tilde{\Omega}_i\of{t,\rv}$ are given by the pulsed Gauss-Laguerre beams of  Eq.~\eqref{eq:rabipp}.

Dissipative processes are not an issue for large detunings, why we can resort to the solution of the Schr\"odinger equation for $t> t_i$  and $\ket{\psi}\equiv \ket{\psi''}$
\begin{equation}
	\label{eq:psitimeevolve}
\ket{\psi(t)} =G(t,t_i)\ket{\psi(t_i)} ,
\end{equation}
with the propagator $G(t,t_i)$ \eqref{eq:propagator}.

For the numerical solution of this two-component, (3+1)D problem, we use Fourier methods with symplectic integrators \cite{Yoshida1990} and operator disentangling \cite{Puri2001}. Analytical solutions are examined
for rectangular pulses (Sec.~\ref{sec:constantRabi})
and the hyperbolic secant pulse (Sec.~\ref{sec:DK}).

\subsection{Ideal Bragg beamsplitter and mirror}\label{sec:idealbs}\vsp
The interaction of a two-state system with laser pulses can be understood qualitatively by the ``pulse area'' \cite{McCall1970}
\begin{equation}\label{eq:pulsearea}
\theta(t)=\int_{-\infty}^{t}\text{d}t'\,
\Omega(t'),
\end{equation}
which is rather a phase by dimension.
In the context of ideal Bragg scattering, the two states are the momentum states $\{\ket{-\kl}_x, \ket{\kl}_x\}$. One can visualize the evolution during the action of the Bragg pulse as a motion on the Bloch sphere \cite{Allen1987}.
A symmetrical 50:50 Bragg beamsplitter corresponds to a $\theta=\pi/2$ rotation from the south pole to the equator at some longitude. This gives equal probability to the outputs channels  
$\ket{\pm\kl}$.
A $\theta=\pi$ rotation from the south pole to the north pole reverses the momenta
$\ket{-\kl}\rightarrow \ket{\kl}$ and thus acts like  a mirror.
In the following discussion, we will focus on the mirror configuration as it is most susceptible to aberrations, due to the longer interaction time.

The polar decomposition of the transition amplitude 
\begin{equation}\label{eq:transferfct}
\bra{\vec{k}'}G(t, t_i)\ket{\vec{k}}= \sqrt{\eta_{k'k}}e^{i\phi_{k'k}}
\end{equation}
between initial $\ket{\vec{k}}$ and final $\ket{\vec{k}'}$ momentum states characterizes the diffraction efficiency $0\leq \eta_{k'k} \leq 1$.
For atomic wavepackets, we use the phase sensitive fidelity
\begin{equation}
\label{eq:fidelity}
F=|\langle \psi_\text{ideal}|\psi(t_f)\rangle|^2, \qquad \ket{\psi_\text{ideal}} = e^{2i\klvec \op{\vec{x}}} |\psi_i\rangle, 
\end{equation}
characterizing the overlap of the final state $\ket{\psi(t_f)}$ of Eq.\,\eqref{eq:psitimeevolve} and the ideal final state $\ket{\psi_\text{ideal}}$.
For an initial plane wave, the fidelity is $F=\eta_{k'k}$ with $\vec{k}'=\vec{k}+2\klvec$.

\subsection{Sources of aberrations}\label{sec:sourcesaberrations}\vsp
The velocity dispersion of Bragg diffraction \cite{Szigeti2012} is significant 
 and leads to  incomplete population transfer atomic wavepackets
(cf. \fig\ref{fig:braggdiffskizze}, Sec.~\ref{sec:velocdisp}).
Another cause for population loss is off-resonant coupling to higher diffraction orders
(cf. Sec.~\ref{sec:higherdifforders}).
This signals the crossover from the deep-Bragg towards the Raman-Nath regime, 
referred to as quasi-Bragg regime \cite{Muller2008}. 

In general, smooth time-dependent laser pulses (cf. Sec.~\ref{sec:temp}) lead to equally smooth beamsplitter responses  (cf. Sec.\,\ref{sec:DK} and \ref{sec:comparetemp}). 
In contrast, smooth spatial envelopes lead to aberrations (cf. Sec.\,\ref{sec:spatial}).
Every Gauss-Laguerre beam exhibits spatial inhomogeneity and wavefront curvature. 
This is relevant for atomic clouds that are comparable in size to the laser beam waist, or for clouds displaced from the symmetry axis. 
Static laser misalignment further degrades the diffraction efficiency.

There are sundry other dynamical sources of aberrations, such as mechanical vibrations of optical elements or stochastic laser noise \cite{Sturm2014}.
The fundamental process of spontaneous emission leads to decoherence and aberrations, too. Fortunately, this can be suppressed by 
 a detuning $|\Delta|\gg \Gamma$ much larger than the linewidth $\Gamma$, as well as  limiting the interaction time.

\section{Plane-wave approximation}\label{sec:pw}\vsp
The basic mechanism of Bragg beamsplitters arises from the 
momentum transfer of plane waves with a real, constant Rabi frequency
$ \tOmega_1(t,\rv)=\tOmega_2(t,\rv)=\Omega_0$ within the duration of a rectangular pulse. This model is the reference to gauge more realistic calculations. 
Consequently,  the  two components 
$\{\psi_e(t,\rv),\psi_g(t,\rv)\}$ of the Schrödinger field evolve according to
\begin{subequations}\label{eq:dglpsi}
	\begin{align}
	&i\partial_t\psi_e=
	\left(-\frac{\hbar}{2\mass}\nabla^2-\Delta\right)\psi_e +\Omega_0\cos(\kl x)\psi_g,\\
	& i\partial_t\psi_g=
	-\frac{\hbar}{2\mass}\nabla^2\psi_g +\Omega_0^\ast \cos(\kl x)\psi_e.
	\end{align}
\end{subequations}
using the Hamilton operator \eqref{eq:Gham}. Assuming the excited state is initially empty,  the atom's kinetic energy is small and the lasers are far-detuned $|\Delta|\gg \Gamma,\ \Omega_0,\  \wrec$, we can adiabatically eliminate the excited state \cite{Brion2007,Muller2008}
\begin{equation}
\psi_e \approx \frac{ \Omega_0}{\Delta} \cos(\kl x)\psi_g.
\end{equation}
Then, the ground state Schr\"odinger equation reads 
\begin{equation}
\label{eq:mathieu}
i\partial _t\psi_g
= \left(-\frac{\hbar}{2\mass}\vec{\nabla}^2
+\mathcal{V}(x)\right) \psi_g,
\end{equation}
with the dipole potential 
$\mathcal{V}(x)=\cos^2\of{\kl x}|\Omega_0|^2/\Delta$ \cite{Marksteiner1995}. 
Stationary solutions of the one-dimensional problem are Mathieu functions \cite{Olver2010}.
Our goal is to formulate a suitable ansatz
 for the (3+1) dimensional non-separable equation with time-dependent pulses. 
\subsection{Bloch-wave ansatz}\label{sec:bloch}\vsp
The Bloch picture is suitable for describing the velocity selective atomic diffraction by a standing laser wave \cite{Kazantsev1990, Wilkens1991,Champenois2001}.
The characteristic translation invariance  of the Hamilton operator \eqref{eq:mathieu}
 by a displacement  of $a_x=\lambda_L/2$ defines  a natural length scale.  Its reciprocal is the  lattice vector $\mathfrak{q}_x=2\pi/a_x=2\kl$.
 It is convenient to embed  the total three-dimensional wavefunction in an orthorohmbic volume with  lengths $(N_x a_x,a_y,a_z)$, with $N_x \in \mathbb{N}$ and to impose periodic boundary conditions $\psi_g(x+N_x a_x,y+a_y,z+a_z)=\psi_g(x,y,z)$. Bragg scattering involves at least two photons, one from each of the counterpropagating lasers. Therefore, the two-photon recoil frequency 
 $\wrecd$ \eqref{eq:recoilfrequencies}
emerges as the frequency  scale.
In terms of the dimensionless length 
$\xi=\mathfrak{q}_x x$ 
and time $\tau=\wrecd t$, the Schr\"odinger field
\begin{equation}\label{eq:psiphi}
\psi_g(t,\vec{r})=
\sum_{r=-\floor{\frac{N_y}{2}}}^{\ceil{\frac{N_y}{2}}-1}
\sum_{s=-\floor{\frac{N_z}{2}}}^{\ceil{\frac{N_z}{2}}-1}
e^{i(r \mathfrak{q}_y y+s \mathfrak{q}_z z-\bar{\omega}_{r,s}\tau)}h^{(r,s)}(\tau,\xi),
\end{equation}
factorizes into one-dimensional fields 
$h^{(r,s)}(\tau,\xi)$ and two-dimensional plane waves with the transversal lattice vectors $\mathfrak{q}_{y,z}=2\pi /a_{y,z}$.
The integers $N_{y,z}\in \mathbb{N}$ define the maximal momentum resolution
$\mathfrak{q}_i^{\text{max}}=\mathfrak{q}_i \floor{N_i/2}$.
Please note the use of the Gauss brackets rounding towards the nearest integer at the ``floor'' $\floor{\,} $ or the ``ceiling'' $\ceil{\,}$. 
With a detuning dependent shift of the frequency, introducing the two-photon Rabi frequency $\omph$,
\begin{align}\label{eq:omegaquer}
\bar{\omega}_{r,s}&=\hbar\frac{r^2\mathfrak{q}_y^2+s^2\mathfrak{q}_z^2}{2\mass \wrecd}+\omph,&
\omph&=\frac{\Omega_r}{\wrecd}=\frac{|\Omega_0|^2}{2\wrecd\Delta},
\end{align}
the Schr\"odinger equation  for each amplitude simplifies to
\begin{equation}\label{eq:1dsgbloch}
i \partial_\tau h(\tau,\xi)
= \left(-\partial^2_\xi+\omph\cos{\xi}\right)h(\tau,\xi).
\end{equation}
By construction, the potential is $2\pi$-periodic and  the eigenfunctions 
$h(\tau,\xi)=e^{-i \tau \omega^{(b)}(q)} 
h^{(b)}(\xi,q)$ are given by 
Bloch-waves $h^{(b)}(\xi,q)$ \cite{Kohn1959, Callaway1991, Grupp2007,Sturm2017} with the lattice periodic function $g^{(b)}(\xi,q)$ for momentum $q$ and band index~$b$
\begin{align} \label{eq:blochhg}
h^{(b)}(\xi,q)&=e^{i q \xi} g^{(b)}(\xi,q), \\
g^{(b)}(\xi+2\pi,q)&=g^{(b)}(\xi,q).
\end{align}
 From the periodic boundary conditions for  the wavefunction
$h^{(b)}(\xi+2\pi N_x,q )=h^{(b)}(\xi,q)$,
one obtains a quantization of the wavenumber $q_{n}=n/N_x$ with $n\in \mathbb{Z}$. 
The interval $-1/2\leq q_{n}<1/2$ defines the first Brillouin zone in the reduced zone scheme, whose extent equals the \emph{crystal momentum } $Q=1$. 

Bloch wavefunctions are also periodic in momentum space
$h^{(b)}(\xi,q+Q)=h^{(b)}(\xi,q)$, provided we define 
\begin{equation}
g^{(b)}(\xi,q)=
\sum_{m=-\mathcal{N}}^{\mathcal{N}-1}
e^{i m \xi} g^{(b)}(m+q),
\end{equation} 
by a Fourier series for a maximal diffraction order $\mathcal{N}\in\mathbb{N}$ with boundary condition$g^{(b)}\left(q+\mathcal{N}\right)=g^{(b)}\left(q-\mathcal{N}\right)=0$.
From a superposition of these Bloch waves, one obtains the ansatz
\begin{equation}\label{eq:phiansatz}
h(\tau,\xi)=
\sum_{n=-\floor{\frac{N_x}{2}}}^{\ceil{\frac{N_x}{2}}-1}
\sum_{m=-\mathcal{N}}^{\mathcal{N}-1}
e^{i(m+q_n)\xi}g(\tau,m+q_n),
\end{equation}
for the time-dependent solution of Eq.\,\eqref{eq:1dsgbloch}, compatible with the Bloch theorem and suitable for numerical computation.
This ansatz transforms the partial differential equation into the parametric difference equation
\begin{equation}
\label{eq:gmdiffeff}
i\partial_\tau g_{m}(\tau,q) =(m+q)^2 g_{m}  +\tfrac{\omph}{2} (g_{m+1} + g_{m-1}).
\end{equation}  
The $q$-dependence of the $m^{th}$-order scattering amplitude 
$g_m(\tau,q)\equiv g(\tau,m+q)$ leads to the velocity dispersion of Bragg diffraction.
Assuming Dirichlet boundary conditions, 
one can use a $(2\mathcal{N}-1)$-dimensional representation
$\vec{g}^e=
(g_{-(\mathcal{N}-1)},\ldots,
g_{\mathcal{N}-1}),
$
to study the initial value problem
\begin{equation}\label{eq:discGF}
i \dot{\vec{g}}^e = H^e(q)\vec{g}^e, \qquad
H^e= D^e + L + L^\dagger.
\end{equation}
For the indices $1-\mathcal{N}\leq m\leq \mathcal{N}-1$, the Hamilton matrix $H^e$ is formed by a diagonal matrix $D^e$ and a lower triangular matrix $L$
\begin{equation}\label{eq:Lmatrix}
D^e_{m,n}=(m+q)^2\delta_{m,n}, \qquad
L_{m,n}=\tfrac{\omph}{2}\delta_{m,n+1}.
\end{equation}

\begin{figure}[t] 
	\centering
	\includegraphics[width=\columnwidth]{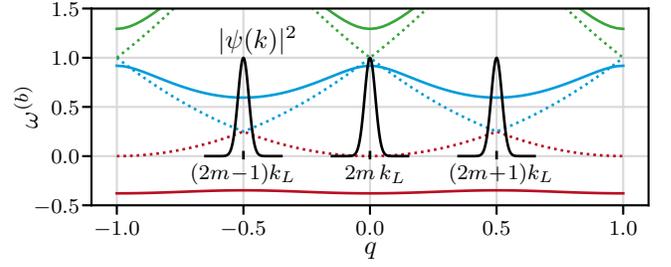}
		\vspace{-6mm}
	\caption{
		Energy bands $\omega^{(0,1,2)}(q)$ of a periodic lattice \mbox{in the} extended zone scheme versus quasi-momentum $q$, with \textit{empty} lattice $\omph=0$ (dotted) and finite depth $\omph=1$ (solid), where $\omphr\!=\!\omph\,\wrecd\!=\!4\,\wrec$. 
		Initial wavepackets with odd  momenta $(2m+1)\kl$ are located at the edges $q\!=\!\pm1/2$  of the $1^{st}$ Brillouin zone, while even momenta 
		$2m\kl$ are at the center $q=0$. 	\label{blochenergiesB}}	
\end{figure}

In order to study the discrete Bloch energy bands $\omega^{(b)}(q)$, one has to solve the eigenvalue problem
\begin{equation}\label{eq:evprob}
\vec{g}^e(\tau,q)=e^{-i\tau\omega(q)}\vec{g}^e(q),\qquad
\omega(q)\vec{g}^e=H^e(q) \vec{g}^e.
\end{equation}
In Fig.~\ref{blochenergiesB}, we present the lowest few energy bands $\omega^{(b)}(q)$ versus the lattice momentum $q$ in an extended momentum zone scheme. For reference, we depict 
the quadratic dispersion relation of the \textit{empty} lattice $\omph\!=\!0$ and the dispersion relation for  $\omph\!=\!1$ ($\omphr\!=\!\omph\,\wrecd\!=\!4\,\wrec$), a moderately deep lattice. 
Narrow  momentum \mbox{wavepackets $\psi(k)$} with
$\sigma_k\ll \kl$ are ideal for beamsplitters.
If they are located at the band edges $k\!=\!q \mathfrak{q}_x\!=\!(\pm1/2+m)2\kl$, the two-photon process covers at least three Brillouin zones. For wavepackets at the center \mbox{$k\!=\!q \mathfrak{q}_x\!=\!2m \kl$,} only two Brillouin zones are coupled by a Bragg pulse.

\subsection{Ad-hoc ansatz}\label{sec:adhoc}\vsp
There are alternatives formulations \cite{Muller2008,Szigeti2012} to the Bloch-wave ansatz, if we define a  Fourier series on the periodic lattice $h(x+N_x a_x)=h(x)$ as 
\begin{equation}\label{eq:phiansatzgeneral}
h(x)
=\sum_{l=-\infty}^{\infty}  e^{i \frac{2\pi l}{N_x a_x} x} g_l, \qquad
\frac{2\pi l}{N_x a_x}=\frac{2 l}{N_x}\kl.
\end{equation}
By decomposing the index $l=N_x m + r$ into a quotient 
$m=\floor{l/N_x}$ and a remainder $0\leq r <N_x$, one obtains
\begin{equation}\label{eq:phiansatzodd}
h(x)=
\sum_{n=-\floor{\frac{N_x}{2}}}^{\ceil{\frac{N_x}{2}}-1}
\sum_{m=-\mathcal{N}}^{\mathcal{N}-1}
g_{2m+1}(\dk_n) e^{i k_{2m+1}^n x} ,
 \end{equation}
with $n=r-\floor{N_x/2}$. In this series, we use a momentum  $k_{\mu}^n =(\mu+\dk_n)\kl$ and a quasimomentum $\dk_n$ 
\begin{equation}
\label{dkodd}
-1 \leq \dk_n=\frac{2n}{N_x}-\frac{\ceil{\frac{N_x}{2}}
	-\floor{\frac{N_x}{2}}}{N_x} < 1,
\end{equation}
in an extended Brillouin zone. 
As the Schr\"odinger equation \eqref{eq:1dsgbloch} has even parity, parity is a conserved quantity. An ansatz with $\sin$ and $\cos$ functions would lead to a  decoupling of \eqref{eq:phiansatzodd} with respect to parity manifolds. 

The decomposition of the index $l=N_x m + n$ is not unique, if we admit signed integral remainders within the limits 
$-\floor{N_x/2}\leq n <\ceil{ N_x/2}$. This implies a quotient
$m=\floor{(l + \floor{N_x/2})/N_x}$. Now, the Fourier series reads
\begin{equation} \label{eq:phiansatzeven}
h(x)=
\sum_{n=-\floor{\frac{N_x}{2}}}^{\ceil{\frac{N_x}{2}}-1}
\sum_{m=-\mathcal{N}}^{\mathcal{N}-1}
 g_{2m}(\dk_n) e^{ik_{2m}^n x},
\end{equation}
with the  quasimomentum $\dk_n$ 
\begin{equation}
\label{dkeven}
-1\leq \dk_n=\frac{2n}{N_x}\leq 1-\frac{1}{N_x}.
\end{equation}
The definitions of the quasimomenta in Eqs.\,\eqref{dkodd} and \eqref{dkeven}, agree exactly for even number $N_x=2u$ of lattice sites or coincide asymptotically for $N_x\rightarrow \infty$. The even/odd ambiguity of number of lattice sites can not be of physical significance as the periodic boundary condition are mere mathematical convenience.  Therefore, assuming an even number of lattice sites is no limitation. 

Using time-dependent amplitudes $g_\mu(\tau,\dk_n)$ in the series \eqref{eq:phiansatzodd} and \eqref{eq:phiansatzeven}, transforms the Schrödinger equation \eqref{eq:1dsgbloch} into a single difference equation $\forall \mu \in\mathbb{Z}$
\begin{equation}
	\label{eq:gndiffeff}
	i \partial_\tau g_{\mu}(\tau,\dk) = \tfrac{1}{4}(\mu+\dk)^2 g_{\mu}+\tfrac{\omph}{2} (g_{\mu+2} + g_{\mu-2}).
\end{equation} 
Due to the two-photon transfer, there is no coupling between even and odd solution manifolds.  Consequently, it is  advantageous to use Eq.\,\eqref{eq:phiansatzodd} for wavepackets located around odd multiples of $\kl$
or Eq.\,\eqref{eq:phiansatzeven} for even multiples of $\kl$ (cf. \fig\ref{blochenergiesB}). As in the comoving frame $S'$ \eqref{comov} mainly $\ket{-\kl}$ is coupled with $\ket{+\kl}$, we focus on the odd solution manifold with $\mu=2m+1$.    
Therefore, Eq.\,\eqref{eq:gndiffeff} can be cast into a tridiagonal system of linear differential equations
\begin{equation}\label{eq:gndiffeffmatrix}
i \dot{\vec{g}}^o = H^o \vec{g}^o,\qquad
H^o= D^o + L + L^\dagger,
\end{equation} for  
$\vec{g}^o=(
g_{-2\mathcal{N}+1},g_{-2\mathcal{N}+3},\ldots
g_{2\mathcal{N}-1})$
with $L$ from \eqref{eq:Lmatrix} and a diagonal matrix
\begin{equation}\label{eq:do}
 D^{o}_{\mu,\nu}=\tfrac{1}{4}(\mu+\dk)^2 \delta_{\mu,\nu} \equiv D_{\mu,\nu} + \varpi\delta_{ \mu,\nu}.
\end{equation}
In the following, it will be prudent to 
adopt a rotating frame 
$\vec{g}^o(\tau)=e^{-i\varpi \tau}\vec{g}(\tau)$
with a 
frequency offset denoted by $\varpi =  (-1+\dk)^2/4$
\begin{gather}
i \dot{\vec{g}} = \mathcal{H}\vec{g}, \qquad
\mathcal{H}= D + L + L^\dagger
\label{eq:rotframeH},\\
D_{\mu,\nu} =\omega_\mu \delta_{ \mu,\nu},
\qquad 
\omega_\mu=\tfrac{1}{4}(\mu+\dk)^2-\varpi.\label{eq:wmu}
\end{gather}
This grounds the frequency $\omega_{-1}=0$.

\section{Aberration analysis}\label{sec:charac}\vsp
Using the ad-hoc ansatz for Bragg scattering, 
we will successively consider more realistic processes to assess their contribution to aberrations. We begin with the plane-wave approximation and consider four temporal Bragg-pulse shapes $f_i(\tau)$. We will analyze their influence on the velocity dispersion as well as losses into higher diffraction orders. Finally, we will add the spatial envelopes of the Gaussian-Laguerre beams  and consider the cumulative effect. 

\subsection{Bragg-pulse shapes}\label{sec:temp}
\begin{figure}[t]
	\centering
	\includegraphics{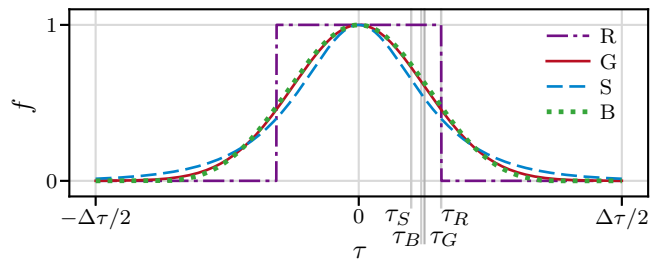}
	\vspace{-8mm}
	\caption{Temporal envelopes $f(\tau)$ for rectangular-,  \mbox{Gaussian-,}  hyperbolic secant- and Blackman pulses for equal nominal time  $T=T_j$, $j\in\{G,R,S,B\}$ and total pulse length $\Delta\tau=8\,\tau_G$. The vertical lines indicates the pulse widths $\tau_j$.}
	\label{fig:tshape}
	\vspace{-2mm}
\end{figure}

\vsp
We examine temporal Gaussian- (G),  rectangular- (R), hyperbolic secant- (S) and Blackman- (B) Rabi pulses 
\begin{equation}\label{rabipulse}
	\Omega(\tau)=\Omega f_j(\tau), \quad j\in\{G, R, S, B\}.
\end{equation}
The shape functions $f_j$, depicted \infig~\ref{fig:tshape}, are all normalized to unity at maximum 
and characterized by a window width $\tau_j$. Different Rabi pulses \eqref{rabipulse} can be compared physically, if they cover the same pulse area \eqref{eq:pulsearea}
\begin{subequations}\label{eq:pulseareaspec}
	\vspace{-4mm}
 	\begin{align}
 		\theta&\equiv\theta(\tau=\infty)=\omph T, \\
 		T&\equiv T(-\infty,\infty),\quad
 		T(\tau_i,\tau_f)=\int_{\tau_i}^{\tau_f}\text{d}\tau\,f_j(\tau),
 		\vspace{-7mm}
 	\end{align}
 \end{subequations}
for equal nominal time $T=T_G=T_R=T_B=T_S$.\\
\textit{Rectangular pulses} are popular in theory as they are constant during the interaction time and lead to simple analytical approximations. They read 
\begin{equation}\label{eq:rec}
f_R(|\tau|\leq \tau_R)=1,\qquad
T_R=2 \tau_R 
\end{equation}
and $f_R(|\tau|>\tau_R)=0$, elsewhere.\\
\textit{Gaussian pulses}
are the standard shapes in pulsed laser experiments
\begin{equation}\label{eq:gauss}
	f_G(\tau) =e^{-\frac{\tau^2}{2\tau^2_G}},\qquad
	T_G=\sqrt{2\pi} \,\tau_G,
\end{equation}
with Gaussian width $\tau_G$.\\
\textit{Blackman pulses}
are characterized by a window function 
\begin{align}\label{blackman}
f_B (\tau) &=w_B\nss\left(\frac{\tau}{\tau_B}\right),\qquad
T_B=\frac{21\pi}{25}\tau_B, \\
w_B(|\phi| \leq \pi)&=\tfrac{1}{50} [21+25\cos(\phi)+4\cos(2\phi)]
\end{align}
and $w_B(|\phi| > \pi)=0$ elsewhere.\\
\textit{Hyperbolic secant pulses} are defined with
\begin{equation}\label{eq:sech}
f_S(\tau)  =\sech\left(\frac{\tau}{\tau_S}\right),\qquad
T_S=\pi \tau_S.
\end{equation} 
They are amenable for analytical solutions \cite{Demkov1969, Vitanov2007}.  

\subsection{Definition of $\pi$- and $\frac{\pi}{2}$-pulses}\vsp
The symmetrical 50:50  beamsplitter pulse and  
the  0:100 mirror pulse are the two most relevant applications of atomic Bragg diffraction
(cf. Sec.~\ref{sec:idealbs}).
Irrespective of the shape, a symmetrical beamsplitter pulse is defined by a pulse area of $\theta=\pi/2$, while a complete specular reflection in momentum space is achieved for  $\theta=\pi$. This defines the nominal times
\begin{equation}\label{eq:nominaltime}
T_{\pi}=\frac{\pi}{|\Omega|},\qquad
T_{\pi/2}=\frac{T_{\pi}}{2}.
\end{equation}
In particular, the four pulse shapes yield mirror widths
\begin{equation}\label{eq:sigmapi0}	
\tau_{G\pi}\!=\!\frac{\sqrt{\pi}}{\sqrt{2}|\omph|},\ 
\tau_{R\pi}\!=\!\frac{\pi}{2|\omph|},\ 
\tau_{B\pi}\!=\!\frac{25}{21 |\omph|},\ 
\tau_{S\pi}\!=\!\frac{1}{|\omph|}.
\end{equation}
Due to the linearity, the symmetric beamsplitter width is just a half of the mirror time i.\,e., $\tau_{\pi/2}=
\tau_\pi/2$.

\subsection{Diffraction efficiency of a rectangular pulse}\label{sec:constantRabi}\vsp
\begin{figure*}[t]
	\includegraphics[width= \textwidth]{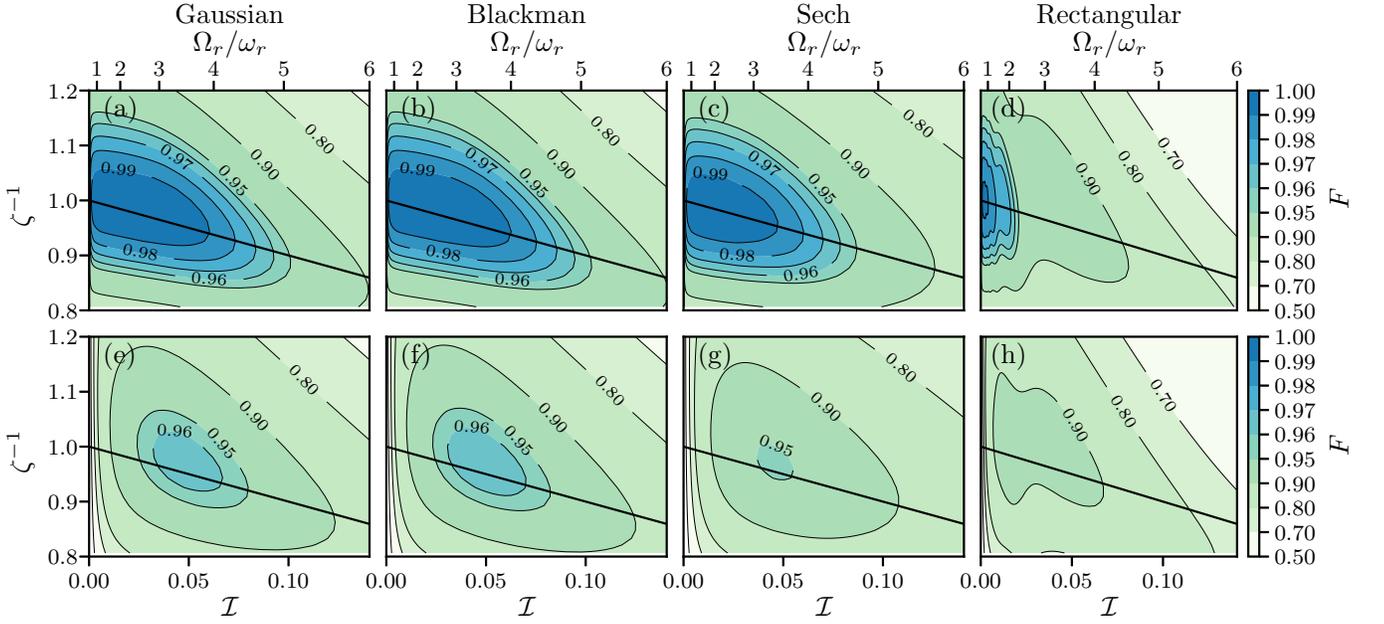}
	\vspace{-6mm}
	\caption{Fidelity $F$ versus two-photon intensity $\bomph=\omph^2/16$, respectively two-photon Rabi frequency $\omph_r=\omph\wrecd$, and inverse $\pi$-pulse stretching factor $\zeta^{-1}=\tau_{j\pi}/\tau_j$, $j\in\{G,B,S,R\}$, for Gaussian (a, e), 
		Blackman (b, f), sech (c, g) and 
		rectangular pulses (d, h). The initial state is a 1D Gaussian wavepacket \eqref{eq:gaussND}, initially centered at $(x, k_x)=(0, -\kl)$ with momentum width $\sigma_k=0.01\, \kl$ (top),  $\sigma_k=0.1\, \kl$ (bottom). 
		The optimal stretching factor $\zeta_\pi$  \eqref{eq:tausigmapi} (solid line) traverses the regions of maximal fidelity. For the numerical (1+1)D integration \eqref{eq:psitimeevolve} with pulse widths $\zeta\tau_{j\pi}$, and total pulse length $\Delta\tau_j=8 \zeta\tau_G$, typical laser and atom parameters, used in experiments (Tab.\,\ref{tab:parameters}), are applied.}
	\label{fig:fidelity_diff_sigma}
\end{figure*}

\subsubsection{Velocity selective Pendellösung} \label{sec:velocdisp}\vsp
In the deep-Bragg regime $\mathcal{N}=1$, 
off-resonant  diffraction orders are negligible.
Thus, for first order diffraction $N=1$ the state vector in the beamsplitter manifold 
\begin{equation}\label{eq:kpm}
k_{\pm}\equiv (\pm1 + \dk) \kl,
\end{equation}
simplifies to the amplitude tuple $\mathbf{g}_\mp(\tau)=(g_{-1},
g_{+1})$ with  $\mathbf{g}_\mp(\tau_i)=(1,0)$. 
The well known Pendellösung \cite{Ewald1917, Zeilinger1986}
\begin{align}\label{eq:pendelsol}
\begin{split}
g_{-1}(\tau)&= 
e^{-i \varphi}
\left(
\cos\vartheta -
\frac{\dk}{i\ompheffdelta} \sin\vartheta
\right),\\
g_{+1}(\tau) &= e^{-i \varphi}
\frac{\omph}{i\ompheffdelta}  \sin\vartheta,
\end{split}
\end{align}
depends on $\varphi=\dk(\tau-\tau_i)/2$, 
$\vartheta=\ompheffdelta(\tau-\tau_i)/2$ 
and the generalized two-photon Rabi frequency $\ompheffdelta= \sqrt{\dk^2 + \omph^2}$. It follows from \eqref{eq:rotframeH} for the rectangular pulse shape \eqref{eq:rec}
\begin{equation}\label{eq:soe}
i\dot{\vec{g}}_\mp(\tau)=\mathcal{H}_{\mp}\vec{g}_\mp, \qquad
\mathcal{H}_{\mp}= 
\begin{pmatrix}0 &\frac{\Omega}{2}\\\frac{\Omega}{2}&\dk\end{pmatrix}.
\end{equation}
With this solution the mirror pulse width \eqref{eq:sigmapi0} can be generalized for arbitrary $\dk\neq0$. 
Maximal efficiency $\eta_{+-}(\tau_{R\pi})=|g_{+ 1}(\tau_\pi)|^2$ is achieved  
 for $\vartheta=\pi/2$, which determines the mirror pulse width
\begin{equation}
	\tau_{R\pi}(\dk)=\frac{\pi}{2 \ompheffdelta}.
\end{equation}
On resonance ($\dk=0$), we recover Eq.\,\eqref{eq:sigmapi0}.
Finally, the diffraction efficiency reads 
\begin{equation}\label{pendeleta}
\eta_{+-}(\tau_{R\pi})=\frac{\Omega^2}{ \ompheffdelta^2}\sin^2{\vartheta_\pi} ,\qquad
\vartheta_\pi = \frac{\pi}{2}\frac{\ompheffdelta}{\omph}.
\end{equation}
The relative phase of the transfer function \eqref{eq:transferfct}, between the final $k_-$ and $k_+$ components is
\begin{equation}\label{eq:phaseshift}
\Delta \phi \equiv \phi_{--}- \phi_{+-}=
\arctan\!\left(\!\frac{\dk}{\ompheffdelta}\tan\vartheta\!\right)\!-\nms\frac{\pi}{2}.
\end{equation}
For $\vartheta =\vartheta_\pi$, one obtains the phase shift after a mirror pulse $\Delta\phi(\tau_{R\pi})$.

\subsubsection{Losses into higher diffraction orders}\label{sec:higherdifforders}\vsp
The transfer function $\bra{\vec{k}'}G (t, t_i)\ket{\vec{k}}$ \eqref{eq:transferfct} exhibits resonances at $\vec{k}^\prime = \vec{k}+2N\klvec$.
On the one hand, resonances with $N\neq 1$ lead to a population loss from the $N = 1$ beamsplitter manifold $\lbrace k_\pm\rbrace$ and reduce the diffraction efficiency.
On the other hand, they  diminish the coupling strength within the beamsplitter manifold. Consequently, this increases
the optimal $\pi$-pulse time $\tspi>\spi$ of a Bragg mirror compared to the prediction of the Pendell\"osung \eqref{eq:sigmapi0}. 
Gochnauer\emph{ et al.}~\cite{Gochnauer2019} have demonstrated this effect experimentally for Gaussian pulses, proving that the effective coupling strength is given by the energy bandgap in the quasimomentum space. 

\paragraph{Renormalized $\pi$-pulse time}
The influence of higher order resonances on the beamsplitter manifold can be calculated perturbatively in terms of the generalized two-photon Rabi frequency $\ompheffdelta$. For $\ompheffdelta\rightarrow 0$ 
all momentum states are doubly degenerate with respect to their energies. We employ  Kato's perturbation theory \cite{Kato1949}, as it can describe the generalized degenerate eigenvalue problem \eqref{eq:katoeigprob}. 
Remarkably, Kato's 1\textsuperscript{st} order perturbation theory coincides with the Pendellösung (cf. App.~\ref{app:kato})

From a third order perturbation calculation $\mathcal{O}(\ompheffdelta^4)$, we find the renormalized Rabi frequency 
\begin{equation}\label{eq:effRabi}
\tomph	=\sqrt{\dk^2(1+2\bomph)^2+\omph^2(1-\bomph)^2}
\stackrel{\bomph\ll 1}{\longrightarrow} \ompheffdelta
\end{equation}
within the beamsplitter manifold using the abbreviation $\bomph=\omph^2/16$. 
For weak dressing $\bomph\ll 1$, it reduces to the  generalized Rabi frequency  of the Pendellösung.
From \eqref{eq:sigmapi0}, one can evaluate the $\pi$-pulse time stretching factor 
\begin{equation}\label{eq:tausigmapi}
\zeta_{\pi}^\dk\!=\!\frac{\tilde{\tau}_{R\pi}}{\tau_{R\pi}}\!=\!\frac{\ompheffdelta}{\tomph},\qquad
\zeta_\pi\equiv
\zeta_{ \pi}^{\dk=0}\! = \frac{1}{1 - \bomph}
\approx\! 1+\bomph.
\end{equation}
Figure~\ref{fig:fidelity_diff_sigma} 
depicts a contour plot of the fidelity $F(\bomph,\zeta)$ \eqref{eq:fidelity} for a Bragg-mirror pulse versus the bare two-photon intensity $\bomph$ and the inverse pulse stretching factor $\zeta^{-1}=\tau_{j\pi}/\tau_j$. This representation uncovers a linear relation.
The numerically calculated fidelity \eqref{eq:fidelity} considers four off-resonant diffraction orders  ($\mathcal{N}=5$).
As initial condition, we consider 1D Gaussian wavepackets \eqref{eq:gaussND} centered at $k_0=-\kl$ with momentum width $\sigma_k$, localized in the center of the laser beams $x_0=0$. Here, in the plane-wave approximation, the results are independent of the expansion size. This size $\sigma_x=(2\sigma_k)^{-1}$ follows from the Heisenberg uncertainty.

Clearly, the $\pi$-pulse stretching factor $\zeta_\pi$ \eqref{eq:tausigmapi} traverses the optimal fidelity regions for all pulse shapes and momentum widths, as a universal rule, motivating the effective $\pi$-pulse widths 
\begin{equation}\label{eq:taupieff}
\tilde{\tau}_{j\pi}=\zeta_\pi\tau_{j\pi}, \quad j\in\{G, R, B, S\},
\end{equation}
with $\tau_{j\pi}$ from Eq.\,\eqref{eq:sigmapi0}.

\begin{figure}[t]
	\includegraphics{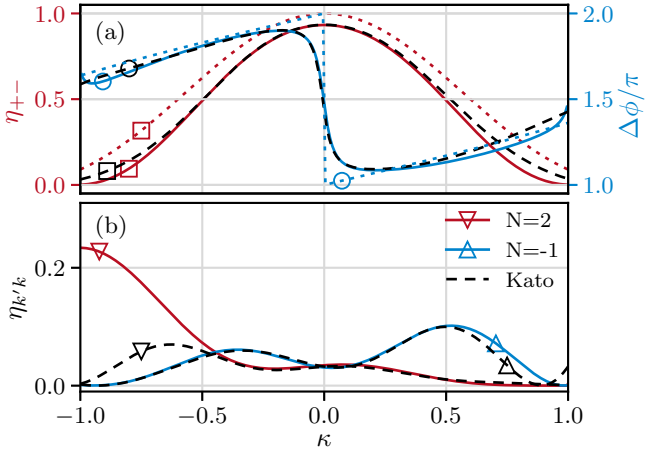}
	\vspace{-6mm}
	\caption{
		(a) Diffraction efficiency $\eta_{+-}$ (\textcolor{tud9c}{\ssquare}) in the beamsplitter manifold $N\!=\!1$, together with the relative phase shift $\Delta \phi$ \eqref{eq:phaseshift} (\textcolor{tud2a}{\scircle}) 		
		and (b) losses into higher diffraction orders $N\!\neq \!1$ versus detuning $\dk$, after a rectangular mirror pulse. 
		For the numerical solution (solid), considering four off-resonant diffraction orders (with $k=(-1+\dk)\kl$ and $k^\prime\!=\! k + 2N\kl$) the applied pulse width is $\tilde{\tau}_{R\pi}(\omph)$ \eqref{eq:taupieff} and for the Pendellösung \eqref{pendeleta}, \eqref{eq:phaseshift} (dotted), considering only the resonant diffraction order, $\tau_{R\pi}(\omph)$ \eqref{eq:sigmapi0} for $\omphr \!=\!\omph\,\wrecd\!=\!\unit[3]{\wrec}$. 
		In (a), the Pendellösung overestimates the efficiency and phase shift, while the Kato corrections \eqref{eq:solkatoform} (dashed) match the numerical results (solid) much better. There are only deviations at the band edges, especially for \mbox{$N\!=\!2$ (b).} }
	\label{fig:velocdispdk}
	\vspace{-2mm}
\end{figure}

\begin{figure}[t]
 \includegraphics{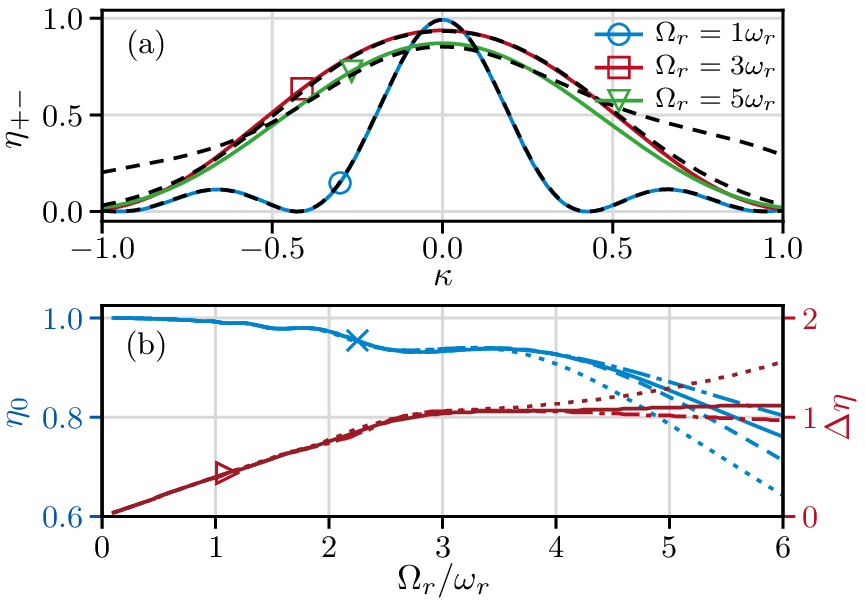}
	\vspace{-6mm}
	\caption{(a) Diffraction efficiency 
		$\eta_{+-}$ after a mirror pulse of width $\tilde{\tau}_{R\pi}(\omph)$ \eqref{eq:taupieff} versus detuning $\dk$ for different two-photon Rabi frequencies \mbox{$\omphr=\omph\,\wrecd$}, numerical results (solid) and Kato \eqref{eq:solkatoform}  solution (dashed). 
		 (b) Resonant transfer efficiency 
		 $\eta_0$ (\textcolor{tud2b}{\scross}) and efficiency width $\Delta\eta$ (\textcolor{tud9c}{\striangright}) 
		versus $\omphr$.
		 The numerically optimal interaction time, for maximal efficiency (dash-dotted) is compared to the approximations for the $\pi$-pulse width $\tau_{R\pi}$ \eqref{eq:sigmapi0} (dotted) and $\tilde{\tau}_{R\pi}$ \eqref{eq:taupieff} (solid). 
		 The  analytical Kato approximation $\eta_0^K(\tilde{\tau}_{R\pi})$ \eqref{eq:dekato} (dashed) provides meaningful predictions.}
	\label{fig:overOm_box}
	\vspace{-3mm}
\end{figure}

\paragraph{Renormalized $\pi$-pulse efficiency}
In \fig \ref{fig:velocdispdk} the velocity dispersion of the response of an atomic mirror is visualized for typical parameters used in experiments (cf. Tab.\,\ref{tab:parameters}) and a two-photon Rabi frequency $\omphr=\omph\wrecd=3\wrec$.
The Pendellösung \eqref{eq:pendelsol}, valid in the deep-Bragg regime ($\mathcal{N}=1$), applying the pulse width $\tau_{R\pi}(\omph)$ \eqref{eq:sigmapi0}, is compared to the eigenvalue solution \eqref{eq:rotframeH} with pulse width $\tilde{\tau}_{R\pi}(\omph)$ \eqref{eq:taupieff}. Therefore, the diffraction efficiency $\eta_{k'k}$ reveals the velocity selectivity of the Bragg condition and the population loss into higher diffraction orders, here in the quasi-Bragg regime ($\mathcal{N}=5$). 
The phase difference $\Delta \phi$ \eqref{eq:phaseshift} shows a $\pi$ jump at resonance. The perturbative Kato solution \eqref{eq:solkatoform}
describes the beamsplitter response very well, only at the band edges $\dk\rightarrow \pm 1$, there are small deviations.
For weak coupling $\omph$, the diffraction efficiency after a mirror pulse of width $\tilde{\tau}_{\pi R}(\omph)$ \eqref{eq:taupieff}, exhibits a sinc behavior [cf. \fig \ref{fig:overOm_box}\,(a)]. It is the typical Fourier-response to a rectangular pulse. 
Increasing the Rabi frequency $\omph$, the response is power broadened, in conjunction with a reduced efficiency. Simultaneously,  the Kato solution becomes less accurate for $|\dk|>0$, while the resonant efficiency $\eta_0\equiv\eta_{+-}(\dk =0)=\eta_{\kl,-\kl}$ can be approximated further. This is also depicted in \mbox{\fig \ref{fig:overOm_box}} (b), together with the efficiency's full width half maximum $\Delta\eta$ of the Bragg mirror. 
For an ideal mirror, $\eta_0=1$ and $\Delta\eta\rightarrow\infty$ are desirable, but impossible. 

In addition, we study the optimal interaction time \infig\ref{fig:overOm_box}\,(b). The approximation $\tau_{R\pi}$ \eqref{eq:sigmapi0} for the deep-Bragg regime and  $\tilde{\tau}_{R\pi}$ \eqref{eq:taupieff} for the quasi-Bragg regime, considering higher diffraction orders, are compared to the optimal interaction time, defined by the maximum numerical transfer efficiency at resonance $\dk=0$. With increasing $\omph$, in a regime where the losses into higher diffraction orders are important, the approximation with $\tau_{R\pi}$ is less accurate, while $\tilde{\tau}_{R\pi}$ can be used further. Please note that for the maximized transfer efficiency the velocity acceptance $\Delta\eta$ is reduced, while for $\tilde{\tau}_{R\pi}$ it remains larger, for increasing $ \omph$. 
From the Kato solution \eqref{eq:solkatoform} a simple analytic equation for the diffraction efficiency on resonance, for the effective $\pi$-pulse time  $\tilde{\tau}_{R\pi}$ can be derived (cf. App.\,\ref{app:kato}) to
\begin{equation}\label{eq:dekato}
\eta_0^K(\tilde{\tau}_{R\pi})=
(1-2\bomph)\left[1+|\omph|\bomph\sin\left(
\frac{2 \pi}{|\omph|}\frac{1+2\bomph}{1-\bomph}\right)\right],
\end{equation} 
also depicted \infig\ref{fig:overOm_box}\,(b). 
This expression predicts losses into higher diffraction orders within the convergence radius $\omphr=\omph\,\wrecd<4\,\wrec$ 
($\mathcal{I}<0.0625$), very well. 
The approximation remains positive 
for $\omphr<8\sqrt{2}\,\wrec$ $ (\mathcal{I}=0.5)$.

\subsection{Diffraction efficiency of a sech pulse}\label{sec:DK}\vsp
\subsubsection{Velocity selective Demkov-Kunike Pendell\"osung}\vsp
For hyperbolic secant pulses $\omph(\tau)=\omph f_S(\tau)$ 
\eqref{eq:sech}, 
one can  solve Eq.\,\eqref{eq:soe} also in a closed form
\cite{Demkov1969, Vitanov2007}.
A decoupling of the first order differential equation system with $g_{+1}=2 i\omph(\tau)^{-1}\dot{g}_{-1}$, leads to Hill's 
second order differential equation \cite{Olver2010} 
\begin{equation}
0=\ddot{g}_{-1}
-\biggl(\frac{\dot\Omega(\tau)}{\Omega(\tau)}  -i\dk\biggr)\,
\dot{g}_{- 1}+ \frac{\Omega(\tau)^2}{4} g_{- 1}.
\end{equation}
With the nonlinear map 
$z(\tau)=
[1+\tanh{(\tau/\tau_S)}]/2$, 
the differential equation for 
$\gamma(z) \equiv g_{-1}(\tau)$ emerges as
\begin{equation}
\begin{aligned}
	\label{eq:hypergeometric}
z(1-z)\gamma''
+ \left[c -z(1+a+b)\right]\gamma'
- a b\gamma= 0,
\end{aligned}
\end{equation}
with $a=\omph\tau_S/2$, $b=-a$ and $c=(1+i\dk\tau_S)/2$.
This is the hypergeometric differential equation with solutions 
$f_1={}_2F_1(a,b;c;z), 
f_2=z^{1-c}{} _2F_1(1+a-c,1+b-c;2-c;z)$ 
and Wronski determinant $w=(1-z)^{c-1} z^{-c}$.
Straightforward analysis (cf. App.\,\ref{app:DK}) leads to the
Demkov-Kunike (DK) solution with unitary propagator $G_\mp(\tau,\tau_i)$
\begin{equation} \label{eq:UDK}
\vec{g}_\mp(\tau)=G_\mp(\tau,\tau_i)\vec{g}_\mp(\tau_i),\qquad
G_\mp(\tau_i,\tau_i)=\mathds{1}. 
\end{equation}
 For the initial datum $\vec{g}_\mp(\tau_i)=(1,0)$, one obtains 
 \begin{equation}\label{eq:DKfull}
 	g_{-1}(\tau)= 
 	 [f_1(\tau) f'_{2}(\tau_i)-f_2(\tau) f'_{1}(\tau_i)]/w(\tau_i).
 \end{equation}
For a pulse beginning in the remote past $\tau_i\ll -\tau_{S}$, this simplifies to
\begin{align}
	g_{-1}(\tau)&=\ _2 F_1\left(a,-a;c,z\right),\label{eq:DKsol}\\
	g_{+1}(\tau)&=\tfrac{a}{i c} \sqrt{z(1-z)}\ _2 F_1\left(1-a,1+a;1+c,z\right).\label{eq:DKsolgp}
\end{align}
Now, the diffraction efficiency of a beamsplitter reads
\begin{equation}\label{eq:etaDK0}
\eta_{+-} ^{{DK}}( \dk, \tau)  = |g_{+ 1}(\tau)|^2=1-|g_{-1}(\tau)|^2.
\end{equation}
Furthermore, for very long pulse durations
$\tau_{S}\ll \tau_{f}, |\tau_{i}|$,  the diffraction efficiency simplifies to 
\begin{equation}\label{eq:etaDK}
\eta_{+-} ^{{DK}}(\dk,\Omega,T)  = \sech^2\!\left(\frac{\dk T}{2}\right) \sin^2\!\left(\frac{\omph T}{2}\right)\!,
\end{equation}
 with the nominal time $T$ \eqref{eq:pulseareaspec}.
In order to achieve full
diffraction efficiency $\eta_{0} ^{{DK}}=\eta_{+-} ^{{DK}}(\dk=0)=1$, one should choose the $\pi$-pulse width as $	\tau_{S\pi}=|\omph|^{-1}$,
 in agreement with the pulse area \eqref{eq:sigmapi0}.
Waiting indefinitely long is hardly ever an option \cite{Beckett1956}. 
Therefore, the finite time approximation
 \vspace{-1mm}
\begin{equation}
	\eta_{0} ^{{DK}}(\tau)\approx z  =\tfrac{1}{2}(1+\tanh{\omph\tau})
	 \vspace{-1mm}
\end{equation} 
reveals the exponential convergence past several $\pi$-pulse times $\tau\gg \tau_S$. It requires $\omphr = \omph\,\wrecd< 3\, \wrec$.

\subsubsection{Losses into higher diffraction orders}\vsp
To consider losses into the higher diffraction orders,  we use time-dependent perturbation theory in	Eq.\,\eqref{eq:rotframeH}
 \vspace{-1mm}
\begin{equation}
i\dot{\vec{g}} =\hdkh(\tau)\vec{g}, \qquad 
\hdkh(\tau)=\hdkh_0 (\tau)+ \hdkh_1(\tau).
 \vspace{-1mm}
\end{equation}
The free evolution $\hdkh_0(\tau)$ consist of a direct sum
 \vspace{-1mm}
\begin{align}
\hdkh_0(\tau) &=
\mathcal{H}_\mp(\tau)
\bigoplus_{\substack{\mu=-\mathcal{N}+1\\ \mu\neq 0,1}}^{\mathcal{N}}\omega_{2 \mu-1}
 \vspace{-1mm}
\end{align}
 of the  DK-generator $\mathcal{H}_\mp(\tau)$ \eqref{eq:soe} in the beamsplitter manifold and the unperturbed  energies  $\omega_\mu$ \eqref{eq:wmu} in the higher momentum states.
The perturbation $\hdkh_1(\tau)$ is simply the complement of the complete Hamilton operator.

\begin{figure}[t]
	\centering
	\includegraphics{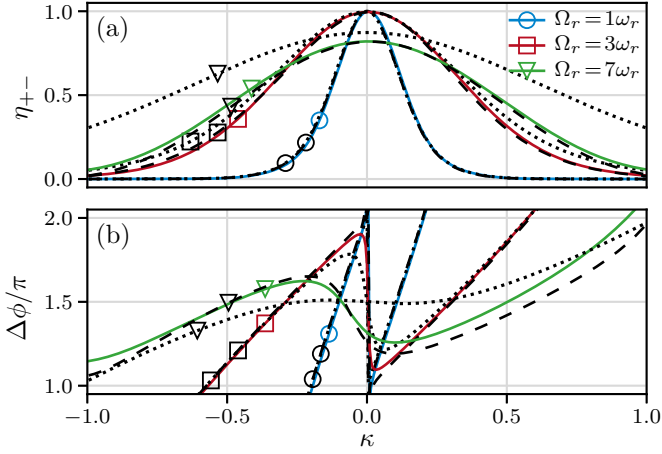}
	\vspace{-8mm}
	\caption{Velocity dispersion of (a) the diffraction efficiency $\eta_{+-}$ and (b) the phase shift $\Delta \phi$  for sech-pulses with  pulse width $\tau_S=\tilde{\tau}_{S\pi}$ \eqref{eq:taupieff} and different Rabi frequencies $\omphr=\omph\,\wrecd$. The DK-Pendellösung \eqref{eq:DKsol} (dotted) is suitable for $\omphr < 3\, \wrec$ while the extended model \eqref{eq:DKsolHO} (dashed) matches the numerical results \eqref{eq:psitimeevolve} (solid) very well also for larger $\omphr$.}
	\label{fig:DK}
\end{figure}

The free retarded propagator is defined for $\tau\ge\tau_i$ as 
 \vspace{-1mm}
\begin{equation}\label{eq:U0}
G_0(\tau,\tau_i) =
G_\mp(\tau,\tau_i)
 \bigoplus_{\substack{\mu=-\mathcal{N}+1\\ \mu\neq 0,1}}^{\mathcal{N}}
e^{-i \omega_{2\mu-1}(\tau-\tau_i)}
 \vspace{-1mm}
\end{equation}
and vanishes elsewhere (cf. App.\,\ref{app:DK}). It involves the DK-Pendellösung $G_\mp$ \eqref{eq:UDK} 
and the  free time evolution of off-resonant momentum states. The complete solution 
 \vspace{-1mm}
\begin{equation}
	\vec{g}(\tau)=G(\tau,\tau_i)\vec{g}(\tau_i)
\end{equation}
follows from the solution $G(\tau,\tau_i)$ of the
integral equation \eqref{DysonSchwinger}.
A second order approximation couples to the $\pm 3\kl,\pm 5 \kl$ momentum states and shifts the frequencies of the beamsplitter manifold
\begin{gather}	\label{eq:DKsolHO}
	G(\tau,\tau_i)=G_0
 -i\int^\infty_{-\infty}\text{d}t\,
 G_0(\tau,t)\mathcal{H}_1(t)
 G_0(t,\tau_i)\\
 -\int^\infty_{-\infty}\text{d}t\text{d}t'\,
 G_0(\tau,t)\mathcal{H}_1(t)
 G_0(t,t')\mathcal{H}_1(t')
 G_0(t',\tau_i).\notag
\end{gather}
This is required to observe the stretching of the $\pi$-pulse time. An explicit analytical approximation can be obtained. It is numerically efficient and useful for the interpretation, but remained unwieldy for display \cite{dissANeumann}.\newline
In \fig\ref{fig:DK}, we compare 
the simple and the extended DK-model after a $\pi$ pulse, with the corresponding numerical (1+1)D simulations \eqref{eq:psitimeevolve}. 
The diffraction efficiency is depicted \infig\ref{fig:DK}\,(a)  and the phase shift $\Delta\phi$ between the coupled states  \infig\ref{fig:DK}\,(b). 
The simple DK-Pendellösung \eqref{eq:DKsol} is valid for  $\omphr\!=\!\omph\,\wrecd \!< \!3\, \wrec$. 
For $\omphr\!>\!3\, \wrec$, losses into higher diffraction orders are significant, but the extended solution \eqref{eq:DKsolHO} still matches the numerical solution. 

\paragraph{Adiabaticity}
The crossover from the deep- to the quasi-Bragg regime at $\omph\approx 3\,\wrec$ for atomic mirrors using $\tilde{\tau}_{j\pi}$ \eqref{eq:taupieff} is related to the adiabaticity criterium \cite{Tong2007}
\begin{equation}\label{eq:adiabaticity}
\max_{\tau\in[\tau_i, \tau_i+\Tint]}\left|\frac{d}{d\tau} \left( 
\frac{\vec{g}_n^o(\tau)^\ast \dot{\vec{g}}_m^o(\tau)}{\omega_n(\tau) -\omega_m(\tau)}\right)\right|\Tint\ll 1,
\end{equation}
$\forall\, m\neq n$, with the eigenvalues $\omega_m(\tau)$ and eigenvectors $\vec{g}_m^o(\tau)$ of $H^o$ \eqref{eq:gndiffeffmatrix}. Equation \eqref{eq:adiabaticity} results in $ \omphr\!=\!\omph\,\wrecd\ll 4\, \wrec$  for $\tilde{\tau}_{S\pi}$ at $\dk=0$. This is confirmed by the results of Gochnauer et al. \cite{Gochnauer2019} and visible in Figs.\,\ref{fig:temp} and \ref{fig:fidelity}.
Therefore, while the DK-Pendellösung \eqref {eq:DKsol} is valid in the adiabatic regime, the extended model \eqref{eq:DKsolHO} can be even used for non-adiabatic pulses.

\subsection{Diffraction efficiency of a Gaussian pulse in the deep-Bragg limit}\label{sec:Gauss}\vsp
Due to the similarity of the Gaussian- to the sech-pulses [cf. Eqs. \eqref{eq:gauss} and \eqref{eq:sech}], one can estimate the velocity selective diffraction efficiency for infinitely long Gaussian pulses in the deep-Bragg regime. The different pulses have equal nominal times \eqref{eq:pulseareaspec}. Therefore, approximating $\sech^2(a)$ from Eq.\,\eqref{eq:etaDK}, with a similar exponential form, providing the same integration area as $\int_{-\infty}^\infty \text{d}a \sech^2(a) =\int_{-\infty}^\infty \text{d}a \exp(-\pi a^2/4)=2 $, leads to
\begin{equation}\label{eq:etaGauss}
\eta_{+-} ^{G}( \dk,\omph, T)  = \exp\left(\!-\pi\Big(\frac{\dk T}{4}\Big)^2\right) \sin^2\left(\frac{T\omph}{2}\right).
\end{equation} 
The results are discussed in the next section.
\subsection{Diffraction efficiency for all pulses in (1+1)D} \label{sec:comparetemp}\vsp
 \begin{figure}[t!]
 	\centering
 	\includegraphics{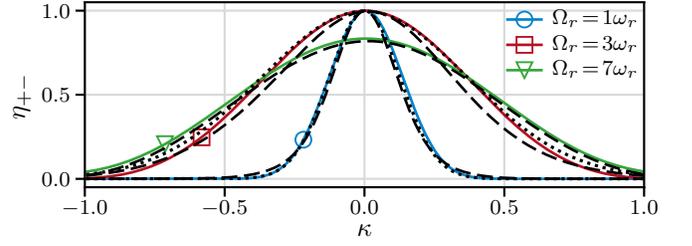}
 	\vspace{-8mm}
 	\caption{Velocity dependent diffraction efficiency $\eta_{+-}(\dk)$ for a Gaussian pulse ($j=G$, solid: numerical, dotted: deep-Bragg limit \eqref{eq:etaGauss}) and the $\sech$ pulse  [$j=S$, dashed: analytical \eqref{eq:DKsolHO}]. A mirror pulse of width $\tilde{\tau}_{j\pi}$ \eqref{eq:taupieff} with total pulse duration $\Tint=8\tilde{\tau}_{G\pi}$ is applied for three Rabi frequencies $\omphr=\omph\,\wrecd$.}
 	\label{fig:diffeffgauss}
 \end{figure}
 \begin{figure}[t!]
 	\centering
 	\includegraphics{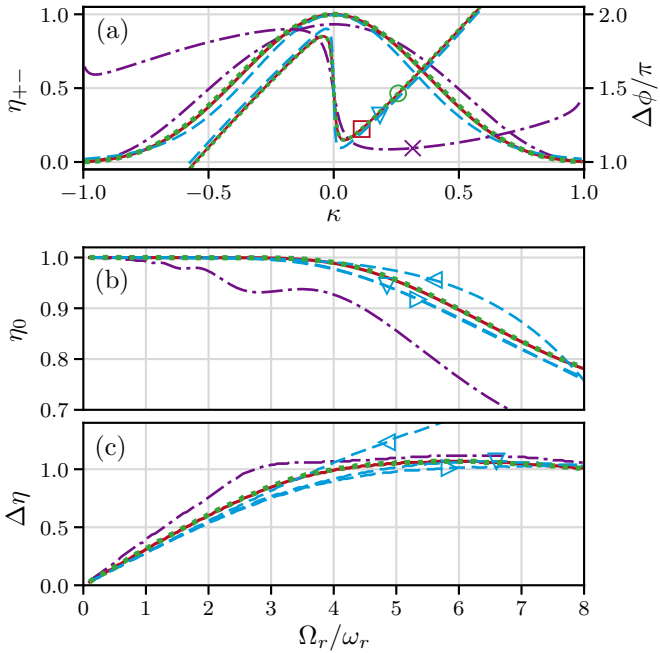}
 	\vspace{-7mm}
 	\caption{Comparison of the Bragg diffraction for a mirror pulse width $\tilde{\tau}_{i\pi}$, for rectangular- (dash-dotted \textcolor{tud11b}{\scross}), Gaussian- (solid \textcolor{tud9c}{\ssquare}), Blackman- (dotted \textcolor{greenb}{\scircle}) and sech-pulses (dashed, numerical: \textcolor{tud2a}{\striangdown}, DK \eqref{eq:DKsol} \textcolor{tud2a}{\striangleft}, DK \eqref{eq:DKsolHO} \textcolor{tud2a}{\striangright}). (a)  Velocity dispersion of the numerical diffraction efficiency $\eta_{+-}$ (without plotmarkers) and phase shift $\Delta\phi$ (with plotmarkers) for \mbox{$\omphr=\omph\,\wrecd=\unit[3]{\wrec}$.} (b) On-resonance diffraction efficiency $\eta_0$ and (c) width of the diffraction efficiency $\Delta\eta$ versus $\omphr$.}
 	\label{fig:temp}
 	\vspace{-2mm}
 \end{figure}
 \begin{figure}[t]
 	\includegraphics{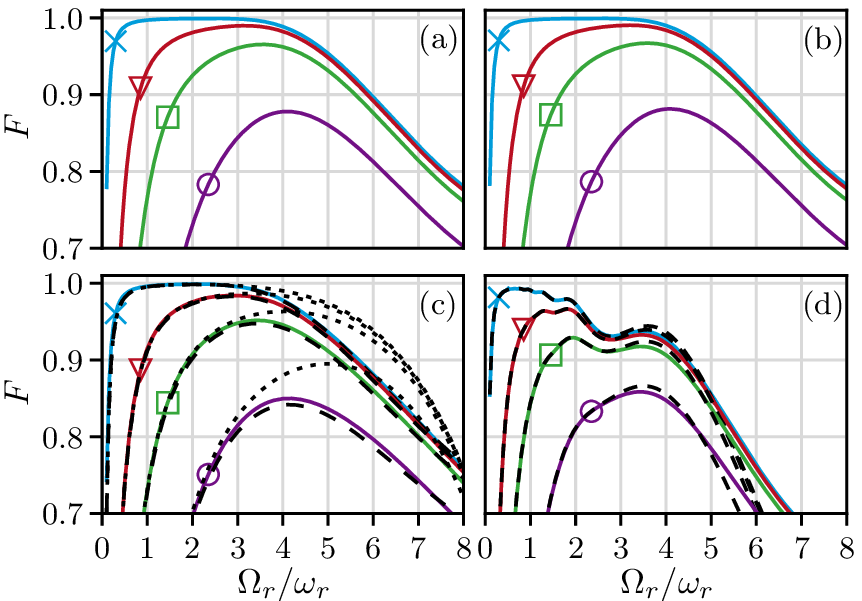}
 	\vspace{-6mm}
 	\caption{Fidelity $F(\omph_r,\sigma_k)$ after a mirror pulse of width $\tilde{\tau}_{i\pi}(\omph)$ \eqref{eq:taupieff} versus the two-photon Rabi frequency $\omphr\!=\!\omph\,\wrecd$ for different initial atomic momentum widths \mbox{$\sigma_k\!=\!\{0.01,0.05,0.1,0.2\}\kl$}, 
 		\{\textcolor{tud2b}{\scross}, \textcolor{tud9c}{\striangdown}, \textcolor{greenb}{\ssquare}, \textcolor{tud11b}{\scircle}\}; for
 		(a) Gaussian, (b) Blackman, (c) sech and (d) rectangular pulses. 
 		The total interaction time is $\Tint=8\,\tilde{\tau}_{G\pi}$  (a-c) 
 		and $\Tint = 2\tilde{\tau}_{R\pi}$ (d), cf. Eq.\,\eqref{eq:taupieff}. 
 		The 1D initial Gaussian wavepacket \eqref{eq:gaussND} is  centered at $(x,k_x)=(0,-\kl)$.
 		The DK-Pendell\"osung \eqref{eq:DKsol} [dotted, (c)] matches the results of the numerical integration \eqref{eq:psitimeevolve} (solid) very well for $\omphr<3\,\wrec$, considering population loss to higher diffraction orders \eqref{eq:DKsolHO} (dashed) also for larger $\omphr$.
 		The Kato solution \eqref{eq:solkatoform} (dashed) is depicted in (d), matching the numerical results. }
 	\label{fig:fidelity}
 \end{figure}
In beamsplitter experiments, Gaussian laser pulses are ubiquitous. 
There is a good reason for it, as they are self-Fourier-transform functions.
This is evident in the numerical simulations of first order diffraction efficiency \infig\ref{fig:diffeffgauss}, which is free of the side lobes of rectangular pulses, seen in \fig\ref{fig:overOm_box}\,(a).
The diffraction efficiency becomes power-broadened for increasing Rabi frequency. Beyond $\omphr>3\omega_r$, scattering into higher diffraction order depletes the population in the beamsplitter manifold. However, in the deep-Bragg regime, the approximation \eqref{eq:etaGauss} matches the numerical solutions very well.
Sech-pulses [extended DK-model \eqref{eq:DKsolHO}] behave similarly, as shown in  \fig\ref{fig:diffeffgauss} and \ref{fig:temp}. 
The explicit solution for the sech-pulse [extended DK-model \eqref{eq:DKsolHO}] deviates
slightly from Gaussian- and Blackman-pulses, but provides very detailed forecasts.
Indeed, all smooth pulse shapes ($j=G,B,S$) with 
pulse widths $\tilde{\tau}_{j\pi}$  are very similar and exhibit
almost identical phase shifts and efficiencies as depicted in \fig\ref{fig:temp}.
Here, for finite total interaction times $\Tint$, the $\pi$-pulse conditions are not met exactly $\omph T_j(-\Tint/2,\Tint/2)\approx\pi$ \eqref{eq:pulseareaspec}.
One could adjust the pulse width $\tilde{\tau}_{j\pi}$ for each pulse shape  $j$ to obtain a $\pi$ pulse individually $\omph T_j(-\Tint/2,\Tint/2)=\pi$, but this leads to unequal nominal times $T_j\neq T$ \eqref{eq:pulseareaspec}
and results in significant phase differences. Thus, we consider the same $\pi$-pulse time $\Tint=8\tilde{\tau}_{G\pi}$ for all pulses and the widths $\tau_j=\tilde{\tau}_{j\pi}$ connected via $T_j=T$, the resulting differences in the pulse areas $\omph T_j(-\Tint/2,\Tint/2)$ \eqref{eq:pulseareaspec} are negligible. 

The phase sensitive fidelity \eqref{eq:fidelity} for different pulse shapes and
momentum widths $\sigma_k$ of an initial Gaussian wavepacket in 1D \eqref{eq:gaussND} are compared in \fig\ref{fig:fidelity}.   
For the smooth envelopes, an increasing $\sigma_k$ reduces the range of admissible Rabi frequencies $\omphr\!= \!\omph\,\wrecd$, which  shifts the optimum to higher values. 
Evidently, the DK-Pendell\"osung \eqref{eq:DKsol} matches numerical simulations for $\omphr < \unit[3]{\wrec}$, while the extended DK-model \eqref{eq:DKsolHO} remains further valid. 
The explicit Kato solution \eqref{eq:solkatoform} matches the results for rectangular pulses very well, demonstrating its applicability for wavepackets with finite momentum width.

\subsection{Diffraction efficiency for spatial Gauss-Laguerre modes with pulse shapes in (3+1)D}
\label{sec:spatial}\vsp
\subsubsection{Gauss-Laguerre modes}
\begin{figure}[t]
	\def\svgwidth{1\columnwidth}  
	\centering
	\includegraphics{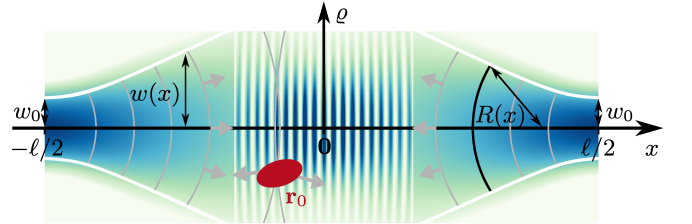}
	\vspace{-4mm}
	\caption{Two counterpropagating, bichromatic Gauss-Laguerre beams form a travelling, standing wave \eqref{vgroup} with an intensity pattern in cylindrical coordinates  $(x,\varrho)$. 
		The  gray arrows are the local wavevectors, $w(x)$ is the local waist and $R(x)$ the local radius of curvature. The distance between the two beam waists is $\ell$. The atomic cloud, generally localized at $\xavec$ is indicated as red ellipse.}
		\label{fig:gaussianbeams}
\end{figure}

The experimental beamsplitter beams are pulsed, bichromatic, counterpropagating Gauss-Laguerre modes \cite{Siegman1986}. In the specific frame $S'$, comoving with the nodes of the interference pattern, there is only a single wavenumber $k_L$ (cf. \eqref{eq:Gham} and Apps.\,\ref{app:galileo}, \ref{app:gaussianbeams}). The slowly varying amplitude of the electric field leads to Rabi frequencies 
\begin{gather}
	\label{eq:spatioGauss}
\Omega_j(t,\rv) =\Omega_{j}(t,\varrho)
e^{
i\Phi(\varrho)},\\
\Omega_{j}(t,\varrho) =\Omega_j(t)
\frac{w_{0}}{w_j}
e^{-\frac{\varrho^2}{w_j^2}},\qquad
\Phi(\varrho)=
\frac{\kl \varrho^2}{2R_j} -\xi_j
\end{gather}
with
beam parameters 
$w_{1,2}=w(\ell/2)$, $R_{1,2}=\pm R(\ell/2)$,
$\xi_{1,2}=\pm\xi(\ell/2)$ and the distance $\ell$ between both lasers beam waists, as depicted in \fig\ref{fig:gaussianbeams}. 
\begin{figure*}[t]
	\centering
	\includegraphics{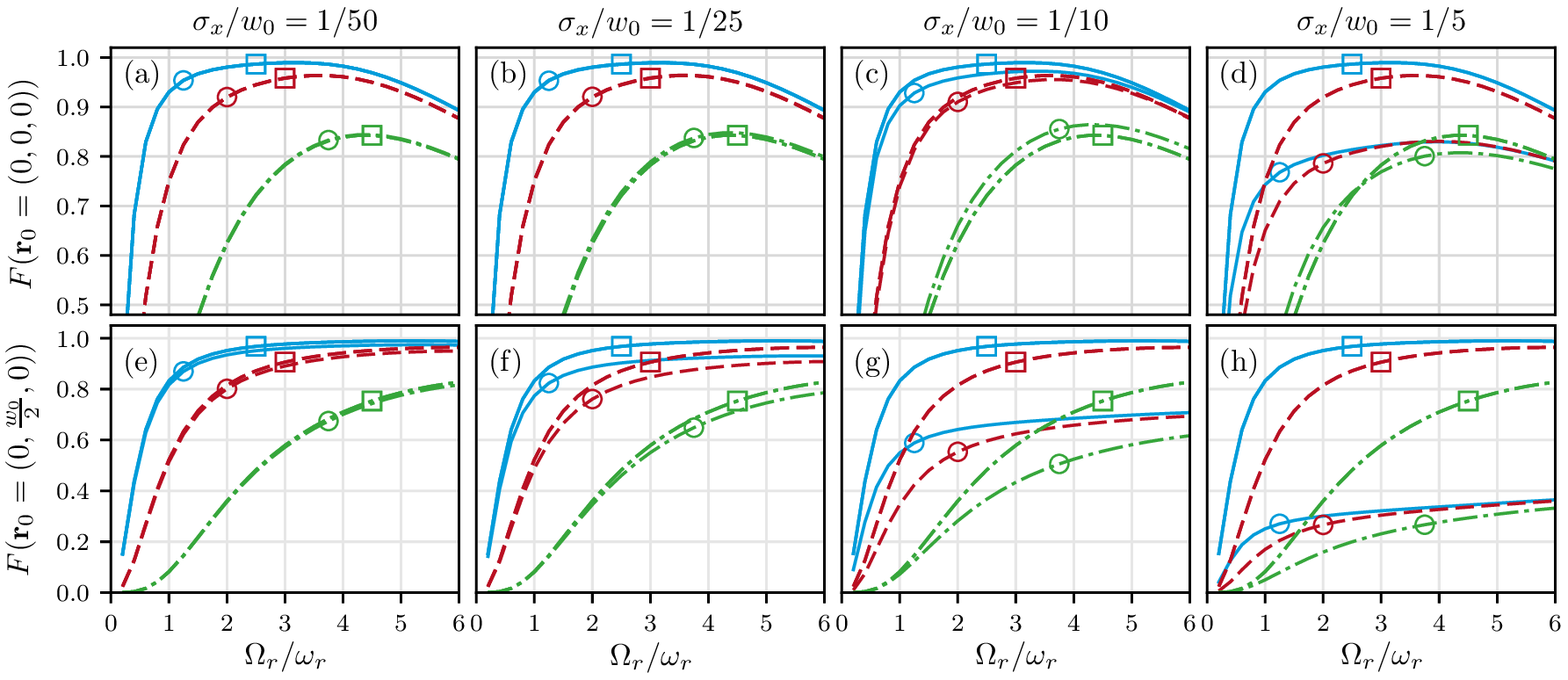}
	\vspace{-8mm}
	\caption{Fidelity $F(\omphr,\sigma_k,\sigma_x)$ after a mirror pulse versus two-photon Rabi frequency $\omphr=\omph\,\wrecd$ for different atomic initial momentum widths \mbox{$\sigma_k=\{0.05,\ 0.1,\ 0.2\}\times\kl$}, \{solid blue, dashed red, dashed-dotted green\} of a 3D ballistically expanded Gaussian wavepacket \eqref{psievolve} for Gauss-Laguerre beams ({\scircle}) in comparison to plane waves ({\ssquare}), using the (3+1)D numerical integration \eqref{eq:psitimeevolve}.  Gaussian temporal pulses of width $\tilde{\tau}_{G\pi}(\omph)$ \eqref{eq:taupieff} and total duration $\Tint=8\,\tilde{\tau}_{G\pi}$ \eqref{eq:taupieff} are applied. Each column represents a different ratio $\sigma_x/w_0$ between spatial width of the initial state $\sigma_x$ and the beam waist $w_0$. In the bottom row the atomic initial state is displaced in the radial direction of the Gaussian beams to $\varrho_0=y_0=w_0/2$.}
	\label{fig:LGPW_Ft}
\end{figure*}
\subsubsection{Local plane-wave approximation}\vsp
To isolate the momentum kick of the beamsplitter from the momentum imparted by the dipole force, we consider a local plane-wave approximation of the Gauss-Laguerre beam at the initial position $\xavec=(0, \vec{\varrho}_0)$, $\vec{\varrho}_0=(y_0, z_0)$ of the atomic cloud
\begin{equation} \label{eq:redrabi}
	\Omega_j(t,\rv) \approx
\Omega_j(t,\xavec)=	\Omega_{j}(t,\varrho_0)	e^{		i\Phi(\varrho_0)}.
\end{equation}
Thus, the atomic cloud feels only a reduced Rabi frequency but experiences no spatial inhomogeneity. Therefore, simulations with plane waves must be independent of the ratio $\sigma_x/w_0$ for $\sigma_x>\lambda$. 

\subsubsection{Simulations}
Beamsplitters perform best, if the atomic cloud (of size $\sigma_x\sim
\si{\micro \meter}-\si{\milli \meter}$) is well localized within the beam 
waist $w_0$.  For $w_0\sim \si{mm}$ and optical wavelengths $\lambda\sim \si{\micro\meter}$ the 
Rayleigh lengths $x_R$ are several meters, thus
\begin{equation}\label{inequ}
x_R \gg w_0>\sigma_x>\lambda.
\end{equation}
Therefore, one can expect that the transversal dipole forces will be stronger than the forces along the propagation direction $x$.
Small clouds centered at the symmetry point $\xavec=(0,0,0)$ will feel the least degradation of the beamsplitter fidelity
[cf. \fig\ref{fig:LGPW_Ft}\,(a) and (b)]
due to dipole forces. 
This will be confirmed by displacing the initial cloud transversely to $\xavec=(0,w_0/2,0)$, leading to larger aberrations [cf.  \fig\ref{fig:LGPW_Ft}\,(e)-(h)].

In these simulations of a Bragg mirror, depicted \infig\,\ref{fig:LGPW_Ft}, we use the effective $\pi$-pulse width $\tilde{\tau}_{G\pi}(\omph(\xavec))$ \eqref{eq:taupieff} in the local plane-wave approximation \eqref{eq:redrabi} for different Rabi frequencies $ \omph_r=\omph\,\wrecd$ and a longitudinal laser displacement $\ell=0.1\,\xr$, like in the experiment (cf. Tab.\,\ref{tab:parameters}). As initial states of the atomic cloud, 
we consider ballistically expanded 3D Gaussian wavepackets \eqref{psievolve} with different widths in 
real space $\sigma_x$ and in reciprocal space $\sigma_k$.

For atoms located at the center of the Gaussian laser beams, the spatial inhomogeneity \eqref{eq:rabipp} leads to significant aberrations only for large atomic clouds [cf. \fig\ref{fig:LGPW_Ft}\,(c), (d)].  
By contrast, even small displaced clouds [cf. \fig\ref{fig:LGPW_Ft}\,(e), (f)] show a significant reduction of the fidelity in \textit{realistic} Gaussian beams compared to ideal plane waves. The latter uses a reduced Rabi frequency according to the local plane-wave approximation \eqref{eq:redrabi}. For large clouds this reduction is detrimental [cf. \fig\ref{fig:LGPW_Ft}\,(g), (h)].  
Please note that we use parameters, where the simulation results for the fidelity only depend on the ratio $\sigma_x/w_0<1$. 

\begin{figure}[t]
	\includegraphics{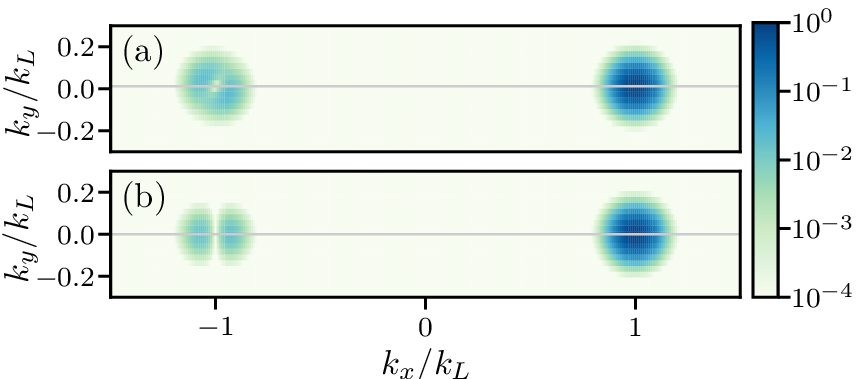}
	\vspace{-6mm}
	\caption{Column integrated atom density 
		 in momentum space $\tilde{n}(k_x,k_y)=
		 \int\text{d}k_z\,
		 |\tilde{\psi}(k_x,k_y,k_z)|^2$ 
		 after a $\pi$ pulse for (a) Gaussian laser beams and (b) plane waves. The initial state is a temporally evolved Gaussian wavepacket \eqref{psievolve} located at $\xavec = (0, w_0/2, 0)$ with momentum width $\sigma_k=0.05\, \kl$  and expansion size $\sigma_x=w_0/25=\SI{2.5}{\micro\meter}$. Gaussian pulses with $\omphr=\omph\,\wrecd=\unit[3]{\wrec}$,  $\tau_G=\tilde{\tau}_{G\pi}(\omph)$ \eqref{eq:taupieff},  $\Tint=\unit[8]{\tilde{\tau}_{G\pi}}$ and beam waist $w_0=\SI{62.5}{\micro\meter}$ are applied. The final momentum expectation value in $y$-direction $\langle \op{p}_{y}\rangle= \unit[0.012] \hbar\kl$  is highlighted with grey lines.}
	\label{fig:density}
	\vspace{-2mm}
\end{figure}

Besides the phase sensitive fidelity, the aberrations due to Gaussian beams are already apparent in the diffraction efficiency. In \fig\ref{fig:density} the momentum density $\tilde{n}(k_x,k_y)$ is shown for the (3+1)D simulation with (a) Gauss-Laguerre laser beams and (b) the idealized local plane-wave approximation, after a mirror pulse with $\omph_r=\unit[3]{\wrec}$. In the momentum space, the splitting is visible directly after the $\pi$-pulse. We study a ballistically expanded Gaussian wavepacket \eqref{psievolve} as initial state with $\sigma_{k}=\unit[0.05]{\kl}$ and  $\sigma_x=1/25\, w_0$, located at 
\mbox{$\xavec = (0, w_0/2, 0)$}. 
The logarithmic scale highlights the imperfections of the Bragg diffraction, using Gaussian laser beams. Even for the tiny momentum width, the diffraction efficiency is reduced to $\unit[96.3]{\%}$ in comparison to $\unit[97.8]{\%}$ for idealized plane waves. In addition, the dipole force leads to a rogue, transversal momentum component $\langle \op{p}_{y}\rangle= \unit[0.012] \hbar\kl$.  As opposed to the diffraction efficiency and the fidelity, this momentum component depends not only on the relation $\sigma_x/w_0$ but on the beam waist, here $w_0=\SI{62.5}{\micro\meter}$. Further studies of the mechanical light effects of the dipole force are subjects of our present research.

Locating the initial state at the center $\xavec = (0,0,0)$ reduces the aberrations due to Gauss-Laguerre laser beams. The diffraction efficiency of $\unit[99.0]{\%}$ reaches almost the efficiency of idealized plane waves with $\unit[99.1]{\%}$ and the transverse momentum component vanishes.

\section{Proving the Demkov-Kunike model experimentally}\label{sec:exp}\vsp
Experimentally, we employ an 
atom chip apparatus  
to Bose-condense $^{87}$Rb \cite{VanZoest2010, Gebbe2019} with a condensate fraction of 
$N^c=\num{10(1)e3}$ and a quantum depletion (thermal cloud) of $N^{\tth}=\num{7(1)e3}$ . 
After release from the trap (lab frame $S$), with trap frequencies listed in Tab.~\ref{tab:parameters}, they expand ballistically
and fall vertically towards Nadir. The Bragg-laser beams are aligned
horizontally. It is sufficient to consider inertial motion during the short Bragg pulses (<\si{\milli s}). 
After $\SI{10}{\milli\second}$ time-of-flight (TOF), at the beginning of the diffraction pulses,   
the temperature of the thermal cloud is obtained from a bimodal fit \cite{Ketterle1999} 
as $T=\SI{20\pm3}{\nano\kelvin}$.
So far, the cloud $\sigma_x =\SI{20}{\micro\meter} $
is much smaller than the beam  waist $w_0 = \SI{1386}{\micro\meter}$ and permits the plane-wave approximation. 

Experimentally, the first order diffraction efficiency in the deep-Bragg limit 
\begin{equation} \label{eq:DE2}
\eta=\frac{N_+}{N_{-}+N_{+}}
\end{equation}
 is obtained from the number of atoms $N_+$ 
diffracted into the first order $k'=k_{+}$ and the undiffracted atoms $N_{-}$  remaining in the initial state $k'=k_{-}$. 
The diffraction efficiency is either a function of the detuning $\df$ \eqref{eq:dw} of the laser from the two-photon resonance with atoms initially at rest $\langle \hat{p}_x(\tau_i)\rangle=0$, or it is the response for resonant lasers and an initial wavepacket centered at 
\begin{equation}\label{eq:pinit}
\langle \hat{p}_x(\tau_i)\rangle=(-1+\bar{\dk})\hbar\kl, \qquad \bar{\dk}=\frac{\df}{\wrecd},
\end{equation}
using Eq.\,\eqref{eq:dwdk} (cf. Sec. \ref{offresponse}).

Theoretically, we compute the diffraction efficiency \eqref{eq:DE2}
in the laser plane-wave approximation
 from the number of diffracted atoms 
\begin{equation}\label{eq:DE3}
\mfn_\pm(\bar{\dk})= \int_{-1}^{1}\text{d}\dk\, \eta_{\pm-} (\dk)\, 
\din(\dk,\bar{\dk}),
\end{equation}
following from a reaction equation derived in App.~\ref{app:DEsigmak}, which completely encloses the wavepacket with the effectively one-dimensional  momentum density $n(\dk,\bar\dk)$  and the average initial momentum $\bar{\dk}$. Please note that for ideal plane matter-waves with wavenumber $\bar\dk$ the diffraction efficiency \eqref{eq:DE2} reduces to $\eta=\eta_{+-} (\bar\dk)$. 
In the deep-Bragg regime, theoretically $\mfn_+ +\mfn_-=N^A=N^c+N^\tth$ and the diffraction efficiency simplifies to
\begin{equation}\label{eq:etadB}
\eta=\frac{\mfn_+(\bar{\dk})}{N^A}=
p^c \gnn_+^{c}(\bar{\dk})+p^\tth \gnn_+^{\tth}(\bar{\dk}),
\end{equation}
splitting into a condensate and a thermal cloud fraction with $p^c=N^c/N^A$, $p^{\tth}=1-p^c$. 
Approximating the normalized initial momentum distributions $\nn^c(\dk,\bar\dk)$, $\nn^{\tth}(\dk,\bar\dk)$ by Gaussian functions \eqref{eq:columndensitygauss} of widths  $\sigma_k^c=0.087\, \kl$ and $\sigma_k^{\tth}=(0.237\pm0.015)\,\kl$ (cf. App. \ref{app:momdist})
and using the Gaussian approximation \eqref{eq:etaGauss} for the diffraction efficiency $\eta_{\pm-} (\dk)$, one obtains the analytical model
\begin{equation}\label{eq:etadfapprox}
\eta= \sin^2\left(\frac{\omph T}{2}\right)
\sum_{a=\{c, {\tth}\}}
\frac{p^a}{\tilde\sigma_k^a(\tilde{T})}
e^{-\frac{(\bar{\dk} \tilde{T})^2}{2\tilde\sigma_k^a(\tilde{T})^2}},
\end{equation}
with $\tilde\sigma_k^a(\tilde{T})=\sqrt{1+(\tilde{T}\tilde\sigma_k^a)^2}$, $\tilde{T}=T\sqrt{\pi/8}$, $\sigma_k^a=\tilde\sigma_k^a\kl$.\\ 

\begin{figure}[t]
	\includegraphics[width=\columnwidth]{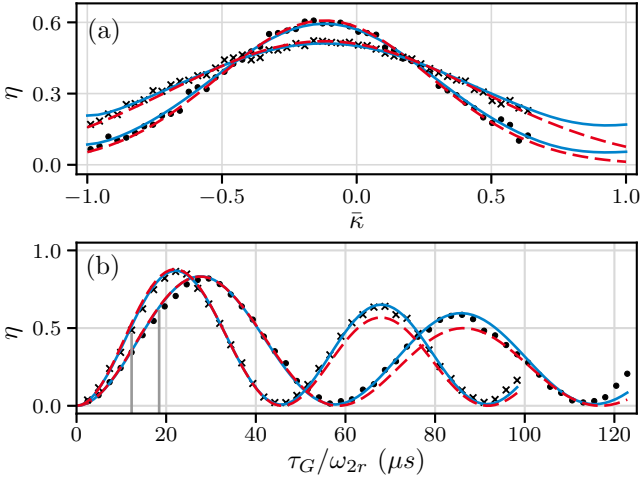}
	\vspace{-8mm}
	\caption{Experimental diffraction efficiency $\eta$ \eqref{eq:DE2} 
		for different laser powers
		$P_\bullet=\unit[20]{mW}$ and $P_{\crosss}=\unit[30]{mW}$ of Gaussian pulses of width $\tau_G$ with numerical simulations (solid, blue) and fits \eqref{eq:etadfapprox} (dashed, red) based on the DK-model.
		(a) Velocity selectivity for $\omph T_G=\unit[0.56]{\pi}$ pulses \eqref{eq:pulseareaspec} versus detuning $\bar{\dk}$ of the initial central momentum$\langle \op{p}_{x}(\tau_i)\rangle = (-1+\bar{\dk}^S+\bar{\dk})\hbar\kl$, were $\bar{\dk}^S\kl=0.12\,\kl$ is a small initial velocity of the atoms in the lab frame $S$ and $\bar{\dk}=\df/\wrecd$ \eqref{eq:dwdk}. 
		 (b) Rabi oscillations of the diffraction efficiency versus pulse width $\tau_G$, with total interaction time $\Tint=8\tau_G$ and highlighted pulse widths of (a).
	 	Other parameters 
		  cf.  Tab.\,\ref{tab:exp1D}, \ref{tab:parameters}.}
	\label{fig:exp1D}
	\vspace{-2mm}
\end{figure}
In \fig\ref{fig:exp1D}, the diffraction efficiency \eqref{eq:DE2} 
is depicted for two different laser powers $P_\bullet=\SI{20}{\milli\watt}$, 
$P_{\crosss}=\SI{30}{\milli\watt}$ of a Gaussian pulse of width $\tau_G$ \eqref{eq:gauss} and total interaction time $\Tint=8\tau_G$. In the experiment, the atoms are displaced axially to
$z_0 =\SI{1165(50)}{\micro\meter}=(\num{0.84(4)})w_0$, while $x_0=y_0= \SI{0}{\micro\meter}$. 
This reduces the effective Rabi frequency at the location of the atoms \eqref{eq:redrabi}.

Fits using the model \eqref{eq:etadfapprox} describe the experimental data already very well and 
provide starting parameters [$p_c$, $\omph(\vec{r}_0)$] for the effective 
(1+1)D numerical simulations 
with Gaussian pulses, fully matching the experimental data. 
The experimental, numerical and fit parameters are listed in Tab.\,\ref{tab:exp1D}.

 In \fig\ref{fig:exp1D} (a), the velocity dispersion of the diffraction efficiency 
uncovers an initial motion $k_x^S=\bar{\dk}^S\kl=0.12\,\kl$ of the atomic cloud in the lab frame $S$. 
Considering this in 
$\langle \hat{p}_x(\tau_i)\rangle =(-1 +\bar{\dk}^S+ \bar{\dk})\hbar\kl$ with $\bar{\dk}=\df/\wrecd$
leads to a very good match of the fit model \eqref{eq:etadfapprox}, the simulations and the experimental data. 
 
In \fig\ref{fig:exp1D} (b), the diffraction efficiency displays damped Rabi oscillations versus the pulse width $\tau_G$. This is a typical inhomogeneous line-broadening caused by the momentum widths 
$\sigma_k^c$, $\sigma_k^{th}$, the two-photon detuning $\df=\bar\dk\wrecd\neq0$ and a residual velocity  $\bar{\dk}^S\neq 0$. It is also revealed by the Gaussian approximation \eqref{eq:etadfapprox}. 
The fit results for the two-photon Rabi frequency are also optimal  for the 
numerical simulations matching the experiment within the error level.
\begin{table}[t]
	\caption{Parameters of \fig\ref{fig:exp1D} for the experiment (e), the numerical simulation (n) and the approximation \eqref{eq:etadfapprox} (a). }
	\centering
	\begin{tabular}{l c | l  r  r}
		\hline\hline
		& &  & \small{$P_\bullet\nms=\nms\SI{20(2)}{mW}$}&
		\small{$P_{\crosss} \nms=\nms\SI{30(3)}{mW}$} \\
		\hline
		& e & $p_c$ & $0.59\pm0.08$ & $0.59\pm0.08$\\
		& e & $\omph$ & $\SI{6.60(66)}{\wrec}$ & 	$\SI{9.89(99)}{\wrec}$\\
		& e &$\omph(\xavec)$ & 	$\SI{1.61(27)}{\wrec}$ & 	$\SI{2.41 (40)}{\wrec}$\\
		& e & $\bar\dk^S\kl$ & $(0.12 \pm 0.01)\,\kl$ & $(0.12 \pm 0.01)\,\kl$\\
		\hline
		(a) & e & $\tau_G/\wrecd$ &$\unit[147.45]{\mu s}$ & $\unit[98.3]{\mu s}$ \\
		& e & $\omph(\xavec)T_G$ \eqref{eq:pulseareaspec} & $\unit[0.56]{\pi}$ &$\unit[0.56]{\pi}$ \\
		& a & $p_c$ & $0.59\pm0.06$ & $0.59\pm0.14$\\
		& a &$\omph(\xavec)$ & $(1.74\pm0.01)\,\wrec$ & $(2.27\pm0.01)\,\wrec$\\
		& n &$p_c$ & $0.59$ & $0.59$\\
		& n &$\omph(\xavec)$ & $1.71\,\wrec$ & $\unit[2.28]{\wrec}$\\
		\hline
		(b) & e & $\df/2\pi$ & $-\unit[2]{kHz}$ & $-\unit[2.5]{kHz}$\\
		& a & $p_c$ & $0.55\pm0.03$ & $0.59\pm0.04$\\
		& a &$\omph(\xavec)$ & $(1.81\pm0.01)\,\wrec$ & $(2.30\pm0.01)\,\wrec$\\
		& n & $p_c$ & $0.52$ & $0.52$\\
		& n & $\omph(\xavec)$ & $1.81\,\wrec$ &  $2.30\,\wrec$ \\
		\hline\hline
	\end{tabular}
	\label{tab:exp1D}
\end{table}

It is worth mentioning that the velocity dispersion of the 
efficiency [\fig\ref{fig:exp1D} (a)] is less sensitive to the condensate ratio $p^c$ 
than the Rabi oscillations [\fig\ref{fig:exp1D} (b)]. 
The Gaussian approximation \eqref{eq:etadfapprox} underestimates the second maxima, but the fit of $p^c$ matches the experimental value within its uncertainty.
The numerical simulations predict a condensate ratio at the lower bound of the experimental ratio, still within the uncertainty. 
The reduction of condensate fraction $p^c$ in simulations and Gaussian approximation 
is equivalent to increasing the momentum width of the condensate or thermal cloud.

Thus, the Gaussian approximation \eqref{eq:etadfapprox} of the DK-model gives an unbiased prediction of the experimental data.
It assumes weak two-photon Rabi frequencies $\omphr(\xavec)<3\, \wrec$, justifying the
Pendellösung \eqref{eq:etaDK} 
and small atomic clouds $\sigma_x\ll w_0$ to approximate Gaussian beams by plane-waves.

\section{Conclusion}\label{sec:concl}\vsp
We present (3+1)D simulations and analytical models of a pulsed atomic Bragg beamsplitter.
Thereby, we characterize ubiquitous imperfections, 
like the velocity dispersion and the population losses into higher diffraction orders. 
We study the influence of four common temporal pulses
 (rectangular-, \mbox{Gaussian-,} Blackman- and hyperbolic sech pulse). 
Clearly, the  diffraction efficiency and the fidelity benefit from Fourier-limited, smooth envelopes. Much insight is gained from the analytical Demkov-Kunike model for a hyperbolic secant pulse \eqref{eq:DKsol}. 
It reveals the explicit dependence on the multitude of physical parameters. 
Due to its similarity with a Gaussian pulse, 
the diffraction efficiency \eqref{eq:etaDK} can also be used for it \eqref{eq:etaGauss}.
For a large parameter regime, the model is verified experimentally and matches the velocity dispersion. The extended DK-model \eqref{eq:DKsolHO} matches also losses into higher diffraction orders.

For a rectangular pulse, we have obtained explicit higher order diffraction results from Kato degenerate perturbation theory, which provide insight in the quasi-Bragg regime. 
Due to a renormalization of the effective Rabi frequency in the beamsplitter manifold, 
one finds significant stretching of the optimal $\pi$-pulse time, which has been seen experimentally \cite{Gochnauer2019}. We find this stretching for all considered pulses in the quasi-Bragg regime and assume it is universal.  

Comparing Gauss-Laguerre beams with plane waves reduces the diffraction efficiency and transfer fidelity, in general. The beam inhomogeneity becomes relevant for $\sigma_x > w_0/10$. 
But even for smaller decentered clouds, the fidelity suffers significantly.
Currently, we investigate the aberrations due to laser misalignment and transversal confinement, which will be reported elsewhere. 

\begin{acknowledgments}\vsp
We like to thank Jan Teske for (3+1)D simulation of the initial Bose-Einstein-condensate, Sven Abend and the members of the QUANTUS collaboration for fruitful discussions.
This work is supported by the DLR German Aerospace Center with funds provided by the Federal Ministry for Economic Affairs and Energy (BMWi) under Grant No. 50WM1957.
\end{acknowledgments}
\appendix

\section{Comoving rotating frame} \label{app:galileo}\vsp
In quantum mechanics, a Galilean transformation is represented by the displacement operator \cite{Gottfried2003}
\begin{equation}
\label{eq:G}
\op{G}(t)=
e^{\frac{i}{\hbar} (\pfr\ropv-\rfr \oppv)}=
e^{-\frac{i}{2\hbar}\pfr \rfr }
e^{\frac{i}{\hbar}\pfr \ropv}
e^{-\frac{i}{\hbar}\rfr \popv}
\end{equation}
with a time-dependent  coordinate
$ \rfr=\vec{\mathfrak{r}}_0+\vec{v} t$ and a momentum $\pfr=m \vec{v}$.
It transforms the corresponding Heisenberg operators as
\begin{equation}
\begin{pmatrix}
\ropv'\\
\oppv'
\end{pmatrix}=
\op{G}
\begin{pmatrix}
\ropv\\
\oppv
\end{pmatrix}
\op{G}^\dagger
=
\begin{pmatrix}
\ropv-\vec{\mathfrak{r}}(t)\\
\oppv-\vec{\mathfrak{p}}
\end{pmatrix}.
\end{equation}
In the Schr\"odinger picture, $\op{G}(t)$ transforms the lab frame state 
$\ket{\psi(t)}=
\op{G}(t)\ket{\psi'(t)}$
into the state $\ket{\psi'(t)}$ of the comoving frame.
Evaluating the comoving-frame Hamilton operator
$\op{H}'$
the Schr\"odinger equation reads 
\begin{align} 
\label{eq:sgtrafo}
&i\hbar \partial_t\ket{\psi'}=\op{H}’\ket{\psi'}
=\op{G}^\dag(\op{H}-i\hbar \partial_t)\op{G}
\ket{\psi'},\\
\label{eq:hamtrafo}
&\op{H}'=  \frac{\op{\vec{p}}^2}{2\mass} +  
\hbar \omega_g\op{\sigma}_g + \hbar \omega_e\op{\sigma}_e+  
V(t,\ropv+\rfr).
\end{align} 
In the frame, moving with the group velocity $\vec{v}=v_g\vec{e}_x$
\eqref{vgroup} in the x-direction, 
 the Doppler shifted laser phases 
\begin{align}
 \phi'_1&=\omega_1 t-k_1 (\opx+\mathfrak{x}_0+v_gt)=\omega_L t
 -k_1 (\opx+\mathfrak{x}_0),\\
\phi'_2&=\omega_2 t+k_2 (\opx+\mathfrak{x}_0+v_gt)=\omega_L t+k_2 
(\opx+\mathfrak{x}_0)
\end{align}
oscillate synchronously with
\begin{equation}
\omega_L=\frac{\omega_1+\omega_2}{2}
\left(1-\beta^2\right)\approx \frac{\omega_1+\omega_2}{2}.
\end{equation}
The second order correction in $\beta=v_g/c$ can be neglected safely in our nonrelativistic scenario.

Another local frame transformation 
$\ket{\psi'}=\hat{F}\ket{\psi''}$, eliminates the rapid temporal oscillations and establishes a single spatial period $\lambda=2\pi/k_{L}$ of the optical potential 
\begin{equation}\label{eq:F}
\op{F}(t)=e^{-i\omega_g t -i\wl t\op{\sigma}_{e}+\frac{i}{2}[ 
k_{12}(\opx+\mathfrak{x}_0)-\chi_{12}]\op{\sigma}_z} .
\end{equation}
Now, the transformed Schr\"odinger equation reads
\begin{align} \label{eq:sgtrafoII}
\begin{split}
&i\hbar \partial_t\ket{\psi''}=\op{H}''\ket{\psi''},
\end{split}\\
\begin{split}\label{eq:hamtrafoII}
&\op{H}''
=  \frac{(\opp_x+\tfrac{1}{2}\hbar k_{12}\hat{\sigma}_z)^2}{2M} +
\frac{\opp^2_y+\opp^2_z}{2M}
-\hbar \Delta\op{\sigma}_e \\
&\phantom{+}+ \frac{\hbar}{2}
\hat{\sigma}^\dag 
\left(\tilde\Omega_1(t,\ropv) e^{ik_L \opx} 
 +\tilde\Omega_2(t,\ropv) e^{-ik_L \opx}
\right) +\text{h.c.}
\end{split}
\end{align} 
with a laser detuning $\Delta=\wl-\wa$, 
a common wavenumber $\kl = (k_1 + k_2)/2$ and 
 a  relative wavenumber mismatch $k_{12}= (k_1 - k_2)/2$. 
Global phases of the Rabi frequencies
$\Omega_i(t,\rv)=\tilde{\Omega}_i(t,\rv)
e^{-i\chi_i}$ 
do vanish with the proper gauge
$\chi_{12}=(\chi_1+\chi_2)/2$ 
and the shifted coordinate origin 
$\mathfrak{x}_0=(\chi_1-\chi_2)/2\kl$.

Please note, $k_{12}=(\omega_1-\omega_2)/c\sim \SI{1e-10}{\micro\meter^{-1}}\sim \SI{1e-11}\,\kl$ is tiny in comparison to other relevant momenta. We will consider Bose-Einstein condensates with Thomas-Fermi radii in the trap of a few microns  (cf. Sec.~\ref{sec:exp} and \ref{app:momdist}), the momentum width can be approximated with the Heisenberg width $\Delta k_\text{TF}^\text{H} = 3/2 r_\text{TF} = \SI{0.15}{\micro\meter^{-1}}$,  considering $r_\text{TF}=\SI{10}{\micro\meter}$, while the Rayleigh width gives $\Delta k_\text{TF}^\text{R} = \SI{0.51}{\micro\meter^{-1}}$ 
\cite{Teske2018}.
In our simulations, we consider atomic initial states as Gaussian wavepackets with momentum widths $\sigma_k\in [0.01, 0.05, 0.1, 0.2]\,\kl$, with $\kl\approx\SI{8}{\micro\meter^{-1}}$, to compare  $\Delta k_\text{TF}^\text{R}$ corresponds to $\sigma_k^\text{R} \approx \Delta k_\text{TF}^\text{R}/3  = 0.02\, \kl$.
After release out of the trap the momentum width of the BEC increases. With temperatures $T\leq \SI{20}{\nano\kelvin}$ this gives rise for momentum widths of a thermal cloud $\sigma_k=\sqrt{k_B T M }/\hbar\leq \SI{0.23}\,\kl$.
 Therefore, $k_{12}$ can be \mbox{neglected safely.}

\section{Spreading Gaussian waves}\label{app:gaussianbeams}\vsp
\paragraph{Matter waves} Ballistically spreading Gaussian wavepackets are useful input states to test a beamsplitter. Using different expansion times $t$, one can vary the position  width $\sigma_x$, while keeping the momentum width $\sigma_k$ constant. 
A $n$-dimensional Gaussian unnormalized wavepacket is defined as 
\begin{gather}\label{eq:gaussND}
		\psi_0(\vec{r}) 
		=
		e^{i\kavec(\rv-\ravec)-\frac{1}{2}(\rv-\ravec) (2\Sigma_0)^{-1} (\rv-\ravec)}\\
		=\int\frac{\text{d}^nk}{(2\pi)^\frac{n}{2}}
			e^{i\kv\rv} \sqrt{|2\Sigma_0|}
		e^{-i\kv\ravec-\frac{1}{2}(\kv-\kavec) (2\Sigma_0) (\kv-\kavec) }\notag
\end{gather}
and centered at $(\ravec,\kavec)=(\langle\rv\rangle,\langle-i\nabla\rangle)$.
The wavepacket is normalized to $\int \text{d}^nr\, |\psi_0|^2=\sqrt{|2\pi\Sigma_0|}$ with the covariance matrix $\Sigma_0=\langle(\rv-\rv_0)\otimes(\rv-\rv_0)\rangle$. The three dimensional 
free Schrödinger equation
\begin{equation}
	i\partial_t\psi(t,\rv)=-\frac{\alpha}{2}\Delta_r\psi, \qquad
	\alpha=\frac{\hbar}{M}
\end{equation}
describes the spreading of a matter-wave using the Fourier-transformed field  $\tilde{\psi}_0(\kv)$ implicitly defined in \eqref{eq:gaussND}
\begin{gather}
	\label{psievolve}
		\psi(t,\vec{r}) =
		\int
		\frac{\text{d}^nk}{(2\pi)^\frac{n}{2}}
		 e^{-i t\frac{\alpha}{2} k^2} e^{i\kv\rv}\tilde{\psi}_0(\kv)\\
		=	\mathcal{A}(t) e^{-i \Theta(t)} 
		e^{i\kavec [\rv-\ravec]-\frac{1}{2}[\rv-\ravec(t)] [2\Sigma(t)]^{-1} [\rv-\ravec(t)] }.\notag
\end{gather}
The evolving center position $\ravec(t)$, spreading covariance $\Sigma(t)$, 
dynamical phase $\Theta(t)$ and scale-factor $\mathcal{A}(t)$ read
\begin{align}
	\label{sigmaevolve}
	\Sigma(t)&=\Sigma_0+i t\frac{\alpha}{2},&
	 \ravec(t)&=\ravec+t\alpha\kavec, \\
	 \Theta(t)&=t\frac{\alpha k_0^2}{2}, &
	 \mathcal{A}(t)&=\sqrt{\frac{|\Sigma_0|}{|\Sigma(t)|}}.
\end{align}
In the simulations, we assume an isotropic initial state with
$\Sigma_{ij}=\delta_{ij}\sigma_x^2$ and
\begin{equation}\label{eq:sigmaxt}
\sigma_x(t)=\sigma_x  \sqrt{1+(t/t_H)^2},
\end{equation}
with the Heisenberg time $t_H=2\sigma_x^{2}M/\hbar$. 

\paragraph{Gaussian laser beams}
The scalar mode of a circularly symmetric Gauss-Laguerre beam propagating along the x-direction  follows from the two-dimensional $n=2$ paraxial approximation of the Helmholtz equation
\begin{equation}\label{paraxialSG}
		i\partial_x u(x,\vec{\varrho})=	-\frac{\beta}{2 }\Delta_\varrho u, \qquad
		\beta=\kl^{-1},\quad \vec{\varrho}=(y, z).
\end{equation}
The spatially evolved mode $u (x,\vec{\varrho})$ follows analogously  from  \eqref{psievolve}, \eqref{sigmaevolve}, substituting $(t,\alpha)\leftrightarrow (x,\beta)$
\begin{gather}
u(x,\vec{\varrho})=
\frac{x_R}{iq(x)} e^{i \frac{\kl\varrho^2}{2 q(x)}}=
U e^{i\Phi}\\
U(x,\varrho) =\frac{w_0}{w(x)}
e^{-\frac{\varrho^2}{w(x)^2}}, \quad
\Phi(x,\varrho)=\frac{\kl \varrho^2}{2R(x)} -\xi(x),\notag
\end{gather}
where $\varrho=\sqrt{y^2+z^2}$ is the normal distance to the symmetry axis and $q(x)=x-i \xr$ is the complex beam parameter \cite{Siegman1986}. It is characterized by the Rayleigh range $\xr=\pi w_0^2/\lambda$, the beam waist $w(x)=w_0(1+(x/\xr)^2)^{1/2}$, the minimum waist $w_0=2\sigma$, the radius of wavefront curvature
$R(x)=x (1+(\xr/x)^2)$, the Gouy phase
$\xi= \arctan(x/\xr)$ and the wavelength $\lambda_L=2\pi/\kl$.

We consider two counterpropagating Gaussian laser beams,
which are symmetrically displaced 
with respect to their waists by a distance $\ell$.
Then, the dipole interaction energy 
in the comoving, rotating frame \eqref{eq:hamtrafoII}, reads
\begin{equation}\label{eq:Vdippp}
\op{V}'' = \frac{\hbar}{2}\op{\sigma}^\dagger
\left[  \Omega_1(t,\rv)e^{i\kl x} +\Omega_2(t,\rv)e^{-i\kl x} \right]    + \text{h.c.},
\end{equation}
with pulse amplitudes $\Omega_j(t)$ and spatial envelopes 
\begin{equation}	\label{eq:rabipp}
\Omega_j(t,\rv) =\Omega_j(t)
U(x_j,\varrho)e^{i \Phi(x_j,\varrho)}.
\end{equation}
We use shifted coordinates
$x_{1/2} = \pm(x+v_g t +\ell/2)$ and beam parameters $w_j=w(x_j)$,
$R_j=R(x_j)$ and
$\xi_j= \xi(x_j)$, which are slowly varying for 
$x\ll \xr$.  Beamsplitter pulses are typical short and one can neglect the ballistic displacement 
$v_g t \sim\si{\micro \meter} \ll \ell, \xr$. For small atomic clouds $\sigma_x<w_0/3$, one can approximate $ x_1\approx - x_2\approx\ell/2$.

\section{Degenerate perturbation theory}\label{app:kato}\vsp
To rectify the Pendell\"osung \eqref{eq:pendelsol} with contributions from higher order diffraction, we employ Kato's method for the stationary eigenvalue problem in the presence of degeneracy \cite{Kato1949}.
All eigenvalues of the diagonal part 
$D_0=D(\dk=0)$  of  the Bragg Hamilton operator
  \eqref{eq:rotframeH} are doubly degenerated $1\leq\alpha\leq 2$ on resonance. 
  Therefore, we consider the flow of the eigensystem 
   $\mathcal{H}(\lambda)
   \mathbf{v}_{i,\alpha}(\lambda)=\omega_{i,\alpha}(\lambda)\mathbf{v}_{i,\alpha}(\lambda)$ with
\begin{equation}\label{eq:pertkato}
\mathcal{H}=
D_0+\lambda \mathcal{V}, \qquad
\mathcal{V}=D(\dk)-D_0+ L+L^\dag,
\end{equation}
for $0\leq\lambda\leq1$ in the  degenerate subspace $\mathcal{E}_{i}$. If we denote  the orthonormal eigenvectors of $D_0$ with $\mathbf{v}_{i,\alpha}^{(0)}$ and their eigenvalues $\omega_{i}^{(0)}$, the eigenvectors of the interacting Hamilton operator $\mathcal{H}_i(\lambda)$ restricted to the subspace 
$\mathcal{E}_i$, are $\mathbf{v}_{i,\alpha}(\lambda)=
P_i(\lambda)
\mathbf{v}_{i,\alpha}^{(0)}$. Now, all efforts are put in the perturbative evaluation  of the projection operator $P_i(\lambda)$, which evolves from the  unperturbed projection $P_i^{(0)}$. This results in 
the generalized eigenvalue problem
\begin{gather}
\label{eq:katoeigprob}
\mathcal{H}_i\mathbf{v}_{i,\alpha}^{(0)}
=\omega_{i,\alpha} 
K_i \mathbf{v}_{i,\alpha}^{(0)},\\
\label{eq:katoeigprobHK}
\mathcal{H}_i=P_i^{(0)}\mathcal{H}P_i P_i^{(0)},\quad 
K_i =P_i^{(0)}P_i P_i^{(0)},
\end{gather}
with power series expressions for the operators
\begin{align}
P_i(\lambda)&=P_i^{(0)} +\sum_{n=1}^{\infty}\lambda^n A_i^{(n)} , \label{eq:projection}\\
A_i^{(n)} &= -\sum_{(n)}S_i^{(k_1)}\mathcal{V} S_i^{(k_2)}\mathcal{V} \dots \mathcal{V} S_i^{(k_{n+1})},\\
\mathcal{H} P_i(\lambda)&=\omega^{(0)}_{i}P_i(\lambda) +\sum_{n=1}^{\infty}\lambda^n B_i^{(n)} ,\\
B_i^{(n)} &= \sum_{(n-1)}S_i^{(k_1)}\mathcal{V} S_i^{(k_2)}\mathcal{V} \dots \mathcal{V} S_i^{(k_{n+1})}.
\end{align}
Here $\sum_{(n)}$ denotes a sum over all combinations of integers $k_i\in\mathbb{N}_0$ satisfying 
$k_1+k_2+\ldots +k_{n+1}=n$ and
\vspace{-4mm}
\begin{equation}\vspace{-1mm}
S_i^{(0)}=-P_i^{(0)},\ 
S_i^{(k>0)} =(S_i)^{k},\,
S_i = \frac{\mathds{1}-P_i^{(0)}}{\omega_i^{(0)} 
	\mathds{1} -D_0}.
\end{equation}

It is straight forward to evaluate $\mathcal{H}_i$ and $\mathcal{K}_i$ from \eqref{eq:katoeigprobHK} for the ground-state manifold $i=1$ to order $\mathcal{O}(\lambda^n)$. We find that a third order truncation of the series 
\begin{equation}\label{eq:h1k1}
\begin{aligned}
		\mathcal{H}_1&=
			\begin{pmatrix}
		0&\frac{\omph}{2}\\ 
		\frac{\omph}{2}&\dk
		\end{pmatrix}
		-2\bomph\begin{pmatrix}
		1&  0\\ 
		0& 1
		\end{pmatrix}
		-
		\bomph\begin{pmatrix}
	\dk&\omph\\ 
	\omph&0
	\end{pmatrix}, 	\\
		K_1&=(1-\bomph)\begin{pmatrix}
	1&  0\\ 
	0& 1
	\end{pmatrix}
	-\bomph\begin{pmatrix}
	\dk&  \frac{\omph}{2}\\ 
	\frac{\omph}{2} & -\dk
	\end{pmatrix},\quad 
	\bomph=\frac{\omph^2}{16}
\end{aligned}
\end{equation}
agrees very good with the numerical results. 
The roots of the characteristic equation 
$|\mathcal{H}_1 -(\omega_1-\omega_1^{(0)}) K_1|=0$, determine the corrected eigenfrequencies of the Pendelösung. As the frequency shifts
$\omega_1(\lambda)-\omega_1^{(0)}$,
are already
$\mathcal{O}(\lambda)$,
 it is consistent to use a lower approximation for $K_1$, which leads to better results at the specified order.
In particular, we have evaluated 
$\tilde{\mathcal{H}}_1= K_1^{-1}\mathcal{H}_1$
and Taylor expanded it at the specified order 
\begin{equation}\label{eq:eq:1dsgblochA}
\tilde{\mathcal{H}}_1=\begin{pmatrix}
-\bomph(2+\dk)&  \frac{\omph}{2}(1-\bomph)\\ 
\frac{\omph}{2}(1-\bomph)& \dk-\bomph(2-\dk)
\end{pmatrix}  +\order{\lambda^4}.
\end{equation}
This leads to the succinct expression for the  eigenvalues and -vectors
\begin{gather}
\omega_{1,\pm}=\frac{\dk}{2}-2\bomph
\pm\frac{\tilde{\Omega}}{2},\quad
	\mathbf{v}_{1,\pm}^{(0)}=
	\begin{pmatrix}
		2(\bomph-1)\sqrt{\bomph}\\
		-\tfrac{1}{2}\dk(1+2\bomph) \pm \tomph
	\end{pmatrix},\notag\\
\tilde{\Omega}=
\sqrt{\dk^2(1+2\bomph)^2+\omph^2(1-\bomph)^2}
\end{gather}
in terms of a corrected Rabi frequency $\tomph$ \eqref{eq:effRabi}.

Analogous to that the eigenvalues of the next subspace, coupling $\mu=\pm3$ and representing the most important loss channel, can be calculated from $\tilde{\mathcal{H}}_3=K_3^{-1}\mathcal{H}_3$
\begin{equation}\label{eq:ht3}
\tilde{\mathcal{H}}_3=
\begin{pmatrix}2(1+\bomph)-\dk& 0\\0&2(1+\bomph)+2\dk\end{pmatrix}
 +\order{\lambda^3},
\end{equation}
skipping the $\lambda^3$ terms, which overestimate the losses into $\mu=\pm 3$. Including higher expansion orders would correct this, but we find that the lower expansion \eqref{eq:ht3} is sufficient.
The eigenvalues and -vectors of $\tilde{\mathcal{H}}_3$ are
\begin{equation}
\omega_{3,\pm} = 2(1+ \bomph)+\frac{\dk}{2}\pm\frac{3\dk}{2},\qquad
(\mathbf{v}_{3,+}^{(0)},\mathbf{v}_{3,-}^{(0)})=
\mathds{1}_2.
\end{equation}
With the eigenvectors $\mathbf{v}_ {i,j}=P_i\mathbf{v}_{i,j}^{(0)}$, defined by the projections \eqref{eq:projection}, also expanded up to $\lambda^3$ for \mbox{$\mu=\pm1$ and $\lambda^2$} for $\mu=\pm3$, the time-dependent solution of the Schrö- dinger equation with the Hamiltonian \eqref{eq:pertkato} results in
\begin{align}\label{eq:solkatoform}
 \vec{g}^K(\tau)&=\frac{\tilde{\vec{g}}^K(\tau)}{|\tilde{\vec{g}}^K(\tau)|^2},\\
\tilde{\vec{g}}^K(\tau)&=\sum_{i=\{1,3\}}\sum_{j=\{+, -\}}
 c_{i ,j}e^{-i \omega_ {i,j} (\tau-\tau_i)} \mathbf{v}_{i,j},
\end{align}
where the integration constants $c_{i }^{(j)}$ are defined by the initial condition  $\tilde{\vec{g}}^K(\tau)=(0, 1, 0, 0)$.

The population of the $\mu=1$ state is of special interest, because it defines the diffraction efficiency $\eta_{+-}$. On resonance ($\dk=0$), already $\tilde{\vec{g}}^K(\tau)$ is approximately normalized. Therefore, it can be approximated
\begin{align}
&\eta_{0}^{K}(\tau)
\approx |\left(\tilde{\vec{g}}^K(\tau)\right)_3|^2 \label{eq:dekatot}\\
&=A\left( 1+B\cos[4\tau'(\bomph-1)\sqrt{\bomph}]+C\cos\theta_+ +D\cos\theta_-\right)\!,\nonumber
\end{align}
with $\theta_\pm=2\tau'(1\pm\sqrt{\bomph}+2\bomph -\bomph^{3/2})$, $\tau'=\tau-\tau_i$ and coefficients expanded up to the suited order $\order{\bomph^2}$
\begin{equation}
\begin{aligned}
A&=\frac{1}{2} -\bomph -\frac{\bomph^2}{2}+\order{\bomph^3}, \quad
B=-1 +\order{\bomph^3}, \\
C&= -D =-4\bomph^{3/2}+\mathcal{O}(\bomph^{5/2}).
\end{aligned}
\end{equation}
After the effective $\pi$-pulse time $\tilde{\tau}_{R\pi}= \pi/[2|\omph|(1-\bomph)]$ \eqref{eq:taupieff} the diffraction efficiency \eqref{eq:dekatot} results in Eq.\,\eqref{eq:dekato}.

\section{Demkov-Kunike model}\label{app:DK}\vsp
The retarded Green's function is defined as 
\begin{gather}
	\label{eq:propagator}
	G(\tau,\tau_i) =
	\mathcal{T}
	e^{-i\int^\tau_{\tau_i} \text{d}t\, \mathcal{H}(t)}\theta(\tau-\tau_i),\\
	[i\partial_\tau -\mathcal{H}(\tau)]G(\tau,\tau_i)=
	i\delta(\tau-\tau_i),
\end{gather}
which hold equally for  the free evolution $G_0(\tau,\tau_i)$ by substituting $\mathcal{H}\rightarrow \mathcal{H}_0$. This leads to the Dyson-Schwinger integral equation 
\begin{equation}\label{DysonSchwinger}
	G(\tau,\tau_i)=
	G_0-i \int^\infty_{-\infty}\text{d}t\,
	G_0(\tau,t')\mathcal{H}_1(t')
	G(t',\tau_i),
\end{equation}
which is central to time-dependent perturbation theory.

The two-dimensional Green's function $G_\mp$ of the DK-model can be expressed completely for $\omph,\dk \neq 0$ with the hypergeometric basis functions $f_1, f_2$ form Eq.\,\eqref{eq:hypergeometric}    
\begin{gather}
G_\mp(\tau,\tau_i)=M(z) S(z)  S^{-1}(z_i) M^{-1} (z_i),  \\
	M=\begin{pmatrix}
		1 & 0\\
		0 & \frac{i}{a}\sqrt{z (1-z)}
	\end{pmatrix}, 
\quad
	S=\begin{pmatrix}
	f_1 & f_2 \\
	f_1' & f_2'
\end{pmatrix}.
\end{gather}
In the important case of exact  resonance $\dk=0$, further simplifications are possible and lead to 
\begin{gather}
G_\mp(\tau,\tau_i)=	\begin{pmatrix}
		\cos{\Delta \varphi} & -i \sin{\Delta \varphi} \\
		-i \sin{\Delta \varphi}& \cos{\Delta \varphi}
	\end{pmatrix},\\
\varphi(z)=\omph \tau_S \arcsin{\sqrt{z}},\qquad
\Delta \varphi=\varphi(z)-\varphi(z_i).
\end{gather}  	
The integrals \eqref{DysonSchwinger} can be solved approximately analytically. However, the expressions are bulky, why we forgo showing them \cite{dissANeumann}.

\section{Diffraction efficiency for partially coherent bosonic fields} 
\label{app:DEsigmak}\vsp
The bosonic amplitude $\hat{a}_g(\bk)$ describes the ground state atoms in momentum space and obeys the commutation relation $[\hat{a}_g^{\phantom{\dagger}}(\bk),\hat{a}_g^\dagger(\bq)]=
\delta(\bk-\bq)$. For a Bose-condesed sample, the single-particle density matrix
\begin{equation}
	\rho(\bk,\bq)\equiv\av{\hat{a}_g^\dagger(\bq)\hat{a}_g(\bk)}
	=\rho^c(\bk,\bq)+\rho^\tth(\bk,\bq),
\end{equation}
separates into a condensate $\rho^c(\bk,\bq)=
\alpha^\ast(\bq) \alpha(\bk)$ and a quantum depletion $\rho^\tth(\bk,\bq)$. 
The momentum density
\begin{equation}\label{eq:densities}
	n(\bk)\equiv\rho(\bk,\bk)=N^A\left[p^c \mathfrak{n}^c(\bk)+p^\tth \mathfrak{n}^t(\bk)\right],
\end{equation}
is the observable in a beamsplitter. It is normalized to 
the total number of $N^A=\int_{-\infty}^{\infty}\text{d}^3k\,n(\bk)=N^c+N^\tth$ atoms, the densities  
$\mathfrak{n}^c, \mathfrak{n}^t$ are probability normalized, thus defining
a condensate fraction $p^c= N^c/N^A$ and a thermal fraction $p^{\tth}=N^{\tth}/ N^A$. 
Dynamically, the classical field $\alpha(t)$ obeys the Gross-Pitaevskii equation 
and extensions thereof for $\rho^\tth(t)$ \cite{Akhiezer1981,Proukakis2013,Walser2000}.

During the short beamsplitter pulse ($<\SI{1}{\milli\second}$), only 
single particle dynamics  \eqref{eq:psitimeevolve} are relevant
\begin{equation}\label{eq:propdens}
\rho(\tau) =G(\tau,\tau_i) \rho(\tau_i)G^\dagger(\tau,\tau_i),
\end{equation}
 for the condensate and the thermal cloud.  
In the plane-wave approximation,  the three-dimensional Fourier propagator 
$\gf(\vec{k},\vec{q})=\gf_\parallel \gf_\perp$ \eqref{eq:transferfct}
factorizes into  the transverse propagator 
\begin{equation}
\gf_\perp(\tau,\bk_\perp,\bq_\perp)=e^{-i\frac{\hbar(k_y^2+k_z^2)}{2M}\tau}\delta^{(2)}(\mathbf{k}_\perp-\mathbf{q}_\perp),
\end{equation}
and the longitudinal Greens function in x-direction
\begin{equation}
G_\parallel(\tau,x,\xi)=\sum_{\mu,\nu,n}
\tfrac{\gf_{\mu,\nu}(\dk_n,\tau)}{N_x a_x}
e^{i(k_{\mu}^n x-k_{\nu}^n\xi)},
\end{equation}
using definitions \eqref{eq:phiansatzodd}, \eqref{dkodd}. 
The discrete Greens-matrix $\gf_{\mu,\nu}( \tau,\dk_n)$ satisfies \eqref{eq:gndiffeffmatrix}
with initial condition $\gf_{\mu,\nu}(0,\dk_n)=\delta_{\mu,\nu}$.
In the continuum limit, one uncovers the momentum conservation on a lattice with $k_x=(\mu+\dk)\kl$ and $q_x=(\nu+\dk')\kl$, from the Fourier transformation
\begin{equation}
	\gf_\parallel(\tau,k_x,q_x)=
	\delta(\dk-\dk^\prime) \gf_{\mu,\nu}(\tau,\dk).
\end{equation} 

All observables are along the x-direction. Thus, we average over the transversal directions and introduce the marginal momentum densities at time $\tau$
\begin{equation} \label{eq:columndens}
	n(\tau,k_x)=\int_{-\infty}^\infty \text{d}k_y\text{d}k_z\, n(\tau,\bk).
\end{equation}
We assume that the initial ensemble is well localized around $k_x=(\nu+\dk)\kl$ with $\nu=-1$, and denote the density  by $n_i(\dk)=n(\tau_i,k_x)$. From the propagation equation \eqref{eq:propdens}, one obtains the final density
$n_f(\dk)=n(\tau_f,k_x)$, with $k_x=(\mu+\dk)\kl$ at diffraction order $\mu$ 
\begin{equation}
n_f(\mu,\dk)= |\gf_{\mu,-1}(\tau_f, \dk)|^2 n_i(\dk).
\end{equation}

Now, we can identify the diffraction efficiency as  
$\eta_{+-}(\dk)=|\gf_{1,-1}(\tau_f,\dk)|^2$ 
and   $\eta_{--}(\dk)=|\gf_{-1,-1}(\tau_f,\dk)|^2$.
Thus, for atomic clouds with initial momentum $\langle \hat{p}_x\rangle=(-1+\bar{\dk})\hbar\kl$ \eqref{eq:pinit}, the number of diffracted atoms read
 \begin{equation}
\mfn_\pm(\bar\dk)=\int\limits_{-1}^{1} \text{d}\dk \,\eta_{\pm-}(\dk) n_i(\dk,\bar\dk),
 \end{equation}
which are the observables in $1^{st}$ order diffraction theory. 

\subsection{Initial momentum distribution}\label{app:momdist}
\begin{figure}[tbh]
	\centering
	\vspace{3mm}
	\includegraphics{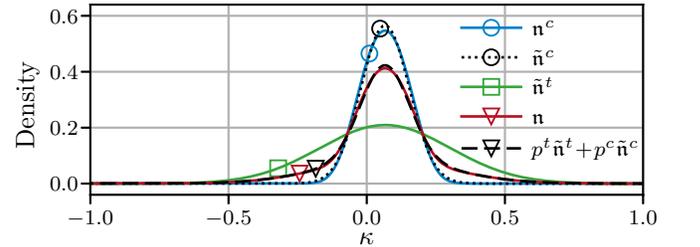}
	\vspace{-8mm}
	\caption{
		One-dimensional density $\mathfrak{n}(\dk)=p^c\mathfrak{n}^c + p^\tth\mathfrak{n}^{\tth}$ \eqref{eq:densities} ($p^t\!=\!0.51, p^c\!=\!0.49$) versus momentum detuning $\dk$. The thermal cloud $\mathfrak{n}^{\tth}$ as well as the condensate $\mathfrak{n}^{c}$ obtained from (3+1)D GP simulation can be approximated with a Gaussian distribution $\mathfrak{n}^{a=\{c,\tth\}}\approx \tilde{\mathfrak{n}}^{a}$ \eqref{eq:columndensitygauss}.}
	\label{fig:initdensity}
\end{figure}

\vsp
After release from the trap, the width of the BEC in momentum space increases due to atomic mean-field interaction \cite{Damon2014}. The momentum distribution is determined by solving the (3+1)D Gross-Pitaevskii equation for the given parameters of Tab.\,\ref{tab:parameters} and $\SI{10}{\milli\second}$ time-of-flight before the diffraction pulses. The result is confirmed by the scaling approach \cite{Castin1996,Kagan1996,Kagan1997a,Meister2017a} applied to the numerical Gross-Pitaevskii ground state.
Finally, the marginal, one-dimensional momentum density distribution of the BEC 
to begin of the diffraction pulses $\mathfrak{n}_i^c\approx\tilde{\mathfrak{n}}^c$ \eqref{eq:columndens}, 
can be approximated with a Gaussian distribution 
\begin{equation}\label{eq:columndensitygauss}
\tilde{\mathfrak{n}}(\dk,\bar{\dk})= \frac{1}{\sqrt{2\pi}\tilde{\sigma}_k}e^{- \frac{(\dk- \bar\dk)^2}{2(\tilde{\sigma}_k)^2}},
\quad \int_{-\infty}^\infty \nms\text{d}\kappa\, \tilde{\mathfrak{n}}(\dk,\bar{\dk}) = 1,
\end{equation}
with the dimensionless momentum width $\tilde\sigma_k=\sigma_k/\kl$ and $\sigma_k^c=0.087\,\kl$, as depicted \infig\ref{fig:initdensity}. 

The thermal cloud is also approximately a Gaussian distribution \cite{Ketterle1999}, where the one-dimensional momentum width $\sigma_k^{\tth} = \sqrt{Mk_B T}/\hbar$ introduces a temperature $T$.
Experimentally, time-of-flight measurements of $\sigma_x(t)$ \eqref{eq:sigmaxt}  
lead to the momentum width $\sigma_k^{\tth}=(0.237\pm0.015)\,\kl$ of $\mathfrak{n}^{\tth}$ \eqref{eq:columndensitygauss} (cf. \fig\ref{fig:initdensity}) and temperature $T=\SI{20\pm3}{\kelvin}$. 
The horizontal trap direction $x'=x\cos \phi$, $\phi=5.5^\circ\pm1^\circ$ differs slightly from the beamsplitter direction $x$. However, the resulting difference in the momentum width $|\sigma_{k_{x}}-\sigma_{k_{x}'}|=0.001\,\kl$ is negligible within the uncertainty. 
\begin{table*}[ht]
	\caption{Experimental parameters: 
		On the $J=\nicefrac{1}{2}\rightarrow J'=\nicefrac{3}{2}$ transition, far-detuned, linearly polarized light couples only to one component of the dipole operator. Therefore, the transition strength is reduced by $\sqrt{3}$. }
	\centering
	\begin{tabular}{l l cc rr r}
		\hline\hline
		\rule{0pt}{10pt} Quantity & & \rule{0pt}{10pt} Symbol & & \rule{0pt}{10pt} Value & & \rule{0pt}{10pt} Reference\\
		\hline\hline
		\multicolumn{7}{c}{\rule{0pt}{10pt}  \textbf{Atom}}\\
		\hline
		Number of atoms in condensate & &$N_c$ & & $\num{10(1)e3}$ && \\
		Number of atoms in thermal cloud & &$N_{\tth}$ & & $\num{7(1)e3}$&& \\
		Atomic mass & & $M$ & & \unit[86.909\,180\,520(15)]{u} && \cite{Bradley1999}\\
		Transition frequency Rb-87 D$_2$ & & $\wa$ & & $2\pi\times384.230\,484\,468\,5(62)\,\si{\tera\hertz}$ &&\cite{Ye1996}\\
		Lifetime & & $\tau$ & & $\SI{26.2348(77)}{\nano\second}$ &&
		\cite{Steck2019}\\
		Decay rate & & $\Gamma$ & & $2\pi\times \SI{6.0666(18)}{\mega\hertz}$&&\\
		D$_2$  ($5^2S_{\nicefrac{1}{2}}\rightarrow 5^2P_{\nicefrac{3}{2}}$) transition dipole matrix element & & $\mathcal{D}$ & & $3.58424(52)\times10^{-29}\,\si{\coulomb\meter}$&&\cite{Steck2019}\\
		Rabi-frequency && $\Omega_0$ && $\mathcal{E}_0 \mathcal{D}/\hbar\sqrt{3}$\\
		Scattering length & & $a$ & & $98.96\, a_0$ &&\cite{Marte2002b}\\  
		Trap frequencies & & $[\omega_{x}, \omega_{y}, \omega_z]$ & & $2\pi \times[46\pm 2, 18\pm1, 31\pm 1]\,\si{\hertz}$&&\\
		Thomas-Fermi radii inside trap & & $[r_{x}, r_{y}, r_z]$ & & $[4.2, 10.8, 6.2] \,\si{\micro\meter}$&&\\
		\hline
		\multicolumn{7}{c}{\rule{0pt}{10pt}\textbf{Laser}}\\
		\hline
		Wavelength & & $\lambda_L$ & & \SI{780.024\,500\,015}{\nano\meter}&&\\
		Wavenumber & & $\kl$ & & $\SI{8.056}{\micro\meter^{-1}}$&& \\
		Detuning to atomic resonance & & $\Delta$ & & \SI{97.875}{\giga\hertz}&&\\
		Beam waist & & $w_0$ & & \SI{1.386}{\milli\meter}  &&\\
		Rayleigh length&  & $\xr$ &  & \SI{7.7}{\meter}&&\\
		Total interaction time & & $\Delta t$ & & $(10^2 ... 10^3)\si{\micro\second}$&&\\
		Gaussian pulse width \eqref{eq:gauss} & & $\tau_G$ & & $\Delta t/8$&& \\
		Distance between laser origins & & $\ell$ & & $ 0.1\,\xr$&& \\
		Total laser power &&	$P$ && $\mathcal{E}_0^2 \epsilon_0 \pi c w_0^2/4$ \\
		Laser amplitude &&  $\mathcal{E}_0$\\
		\hline
		\hline
	\end{tabular}
	\label{tab:parameters}
\end{table*}

\bibliographystyle{apsrev4-1_lessauthors}
\bibliography{library} 
\end{document}